\documentclass[12pt,preprint]{aastex}

\newcommand{\Msun}{~M_\odot}
\newcommand{\lsim}{\raise0.3ex\hbox{$<$}\kern-0.75em{\lower0.65ex\hbox{$\sim$}}}
\newcommand{\gsim}{\raise0.3ex\hbox{$>$}\kern-0.75em{\lower0.65ex\hbox{$\sim$}}}

\newcommand{\cmc}{\rm ~cm^{-3}}
\newcommand{\cms}{\rm ~cm^{-2}}
\newcommand{\kms}{\rm ~km~s^{-1}}
\newcommand{\ergs}{\rm ~erg~s^{-1}}
\newcommand{\wl}{\lambda}
\newcommand{\wll}{\lambda \lambda}
\newcommand{\ml}{~\Msun ~\rm yr^{-1}}
\newcommand{\mll}{\Msun ~\rm yr^{-1}}
\newcommand{\etal}{{et al.\/\ }}

\def\EE#1{\times 10^{#1}}

\begin{document}

\title{OPTICAL AND ULTRAVIOLET SPECTROSCOPY OF SN 1995N:}
\title{EVIDENCE FOR STRONG CIRCUMSTELLAR INTERACTION$^1$}

\author{
Claes Fransson\altaffilmark{2},
Roger A. Chevalier\altaffilmark{3},
Alexei V. Filippenko\altaffilmark{4},
Bruno Leibundgut\altaffilmark{5},
Aaron J. Barth\altaffilmark{6,7},
Robert A. Fesen\altaffilmark{8},
Robert P. Kirshner\altaffilmark{6},
Douglas C. Leonard\altaffilmark{4,9},
Weidong Li\altaffilmark{4},
Peter Lundqvist\altaffilmark{2},
Jesper Sollerman\altaffilmark{5},
 and
Schuyler D. Van Dyk\altaffilmark{10}}

\altaffiltext{1}{Based in part on observations obtained with the {\it
Hubble Space Telescope,} which is operated by AURA, Inc., under NASA
contract NAS 5-26555; in part on observations collected at the
European Southern Observatory, Paranal, Chile; in part on data from
the Lick Observatory, California; and in part on observations from the Keck
Observatory, Hawaii.}
\altaffiltext{2}{Stockholm Observatory, SE--133~36 Saltsj\"obaden,
Sweden.} \altaffiltext{3}{Department of Astronomy, University of
Virginia, P.O. Box 3818, Charlottesville, VA 22903.}
\altaffiltext{4}{Department of Astronomy, University of California,
Berkeley, CA 94720--3411.}
\altaffiltext{5}{European Southern Observatory,
Karl-Schwarzschild-Strasse 2, D-85748 Garching, Germany.}
\altaffiltext{6}{Harvard--Smithsonian Center
for Astrophysics, 60 Garden St., Cambridge, MA 02138.}
\altaffiltext{7}{Present address: Department of Astronomy, 105-24
Caltech, Pasadena, CA 91125.} \altaffiltext{8}{6127 Wilder Laboratory,
Physics \& Astronomy Department, Dartmouth College, Hanover, NH 03755.}
\altaffiltext{9}{Present address: Five College Astronomy Department,
University of Massachusetts, Amherst, MA 01003-9305.}
\altaffiltext{10}{IPAC, 100-22 Caltech, Pasadena, CA 91125.}

\begin{abstract}
Optical and ultraviolet observations of the Type IIn supernova 1995N at
epochs between 321 and 1799 days after the explosion show three distinct
velocity components.  The narrow lines come from circumstellar gas and show
both low and high ionization. This component has a low filling factor, and is
photoionized by X-rays from the shock.  The intermediate component, which is
dominated by newly processed oxygen, originates in a shell with velocity of
2500--5000 $\kms$, and most likely comes from the ejecta. The hydrogen- and
helium-dominated gas has a low ionization, a high density, and 
velocities that extend out to
$\gsim 10,000 \kms$. Strong signatures of Ly$\alpha$-pumped fluorescence lines
of Fe~II are seen in the near-infrared and ultraviolet. The He/H ratio, $\sim
0.3$ by number, and the nitrogen overabundance provide strong evidence for CNO
burning products. The fluxes of the broad hydrogen and helium lines decrease
considerably faster than the oxygen lines. The H$\alpha$ line profile shows
strong evolution, with the red wing decreasing faster than the blue. Possible
scenarios, involving either a clumpy circumstellar medium, or an aspherical
distribution of the surrounding gas, are discussed based on the line profiles
and physical conditions. Finally, we propose that Type IIn supernovae have
their origin in red supergiants in a superwind phase.

\end{abstract}


\section{INTRODUCTION}
\label{sec_introd}

SN 1995N in MCG--02-38-017 (also known as Arp 261) was discovered by Pollas
(1995) on 5 May 1995 (UT dates are used throughout this paper). Its
approximate distance is 24 Mpc, based on $H_0 = 75 \kms$ Mpc$^{-1}$
and the measured recession velocity of the host galaxy (1834 $\kms$;
Theureau et al. 1998).  Optical spectra of the object obtained by
Benetti, Bouchet, \& Schwarz (1995) and by Garnavich \& Challis (1995)
showed it to be a peculiar Type II supernova, with H$\alpha$ emission
(full-width at half-maximum intensity [FWHM] $\approx 1500 \kms$) and narrow
lines of He~I, He~II, [N~II], [O~I], [O~III], [Ne~III], [Ca~V], Fe~II,
[Fe~II], [Fe~III], [Fe~VII], and [Fe~X].  The intensity ratios of the
narrow oxygen lines suggest a high electron density, $> 10^6$
cm$^{-3}$ (Garnavich \& Challis 1995). Benetti et al. (1995) estimated
it to be at least 10 months old, so we will assume an explosion date
of 4 July 1994, befitting for a supernova (in the opinion of the
U.S. coauthors of this paper). Radio emission from the supernova was
discovered at 3.6 cm on 16 June 1995 (Van Dyk et al.  1996). X-ray
emission was discovered with {\it ROSAT} on 23 July 1996, and the
source was subsequently observed with {\it ASCA} on 20 January 1998
(Fox et al. 2000). The unabsorbed X-ray luminosity was $1 \times
10^{41}\ergs$ in the 0.1--10 keV band.

The narrow emission lines observed from SN 1995N place it in the Type
IIn category (Schlegel 1990; Filippenko 1991a,b, 1997), which often
show a slow optical decline compared to other Type II supernovae
(SNe~II).  Baird et al. (1998) have noted the slow evolution of SN
1995N.  The high X-ray and radio luminosities of SN 1995N place it in
a subset of SNe~IIn whose other well-studied members are SN 1986J
(Rupen et al. 1987; Weiler, Panagia, \& Sramek 1990; Leibundgut et
al. 1991; Houck et al. 1998), SN 1988Z (Stathakis \& Sadler 1991;
Filippenko 1991a,b; Turatto et al. 1993; Van Dyk et al. 1993; Fabian
\& Terlevich 1996; Aretxaga et al. 1999),  SN 1978K (Ryder et
al. 1993; Chugai, Danziger, \& Della Valle 1995; Schlegel, Petre, \&
Colbert 1996), and SN 1998S (Leonard et al. 2000, Gerardy et al. 2000;
Fassia et al. 2001).  All of them have X-ray luminosities $\gsim
10^{40}\ergs$, which are at least an order of magnitude higher than
the X-ray luminosities of other SNe~II.  These are also among the most
luminous radio SNe, although SN 1979C does rival them in luminosity,
as does SN 1998bw, a very different kind of object and at a much
earlier epoch (e.g., Galama et al. 1998; Kulkarni et al. 1998).

The early optical observations of SN 1988Z (Stathakis \& Sadler 1991) showed
velocities up to $\sim 20,000 \kms$, and very long baseline interferometry
(VLBI) radio observations of SN 1986J (Bartel, Shapiro, \& Rupen 1989; Bartel
et al. 1991) implied expansion velocities of $\sim 15,000 \kms$.  At the same
time, these SNe showed H$\alpha$ emission lines with FWHM $\approx 500-2000
\kms$, and at least SN 1988Z exhibited very narrow emission lines whose
intensity ratios indicate high gas densities.  This combination suggests that
the rapidly moving supernova material is running into dense circumstellar
clumps and is driving slower shock fronts into them (Chugai 1993; Chugai \&
Danziger 1994; Chugai et al. 1995).  In addition, the combination of high
velocities with strong circumstellar interaction requires that the supernova
have considerable high-velocity gas in order not to be rapidly decelerated.
For SN 1988Z, Chugai \& Danziger (1994) estimate that the total mass in
supernova ejecta is only $\sim 0.5 \Msun$ if the explosion energy is the
standard $10^{51}$ ergs.

In order to more fully understand the nature of this peculiar subset
of SNe~IIn, we have undertaken a detailed spectroscopic study of 
SN 1995N, one of the most important and thoroughly studied examples.
We present the ground-based and {\it Hubble Space Telescope
(HST)} observations in \S~2. In \S~3, we discuss the three main kinematic
components visible in the emission-line profiles. Synthetic spectra are
presented in \S~4, and the temporal evolution of features is examined in
\S~5. We discuss the results and their implications in \S~6, and we summarize
our conclusions in \S~7.

\section{OBSERVATIONS}
\label{sec_obs}

\begin{figure}  
\plotone{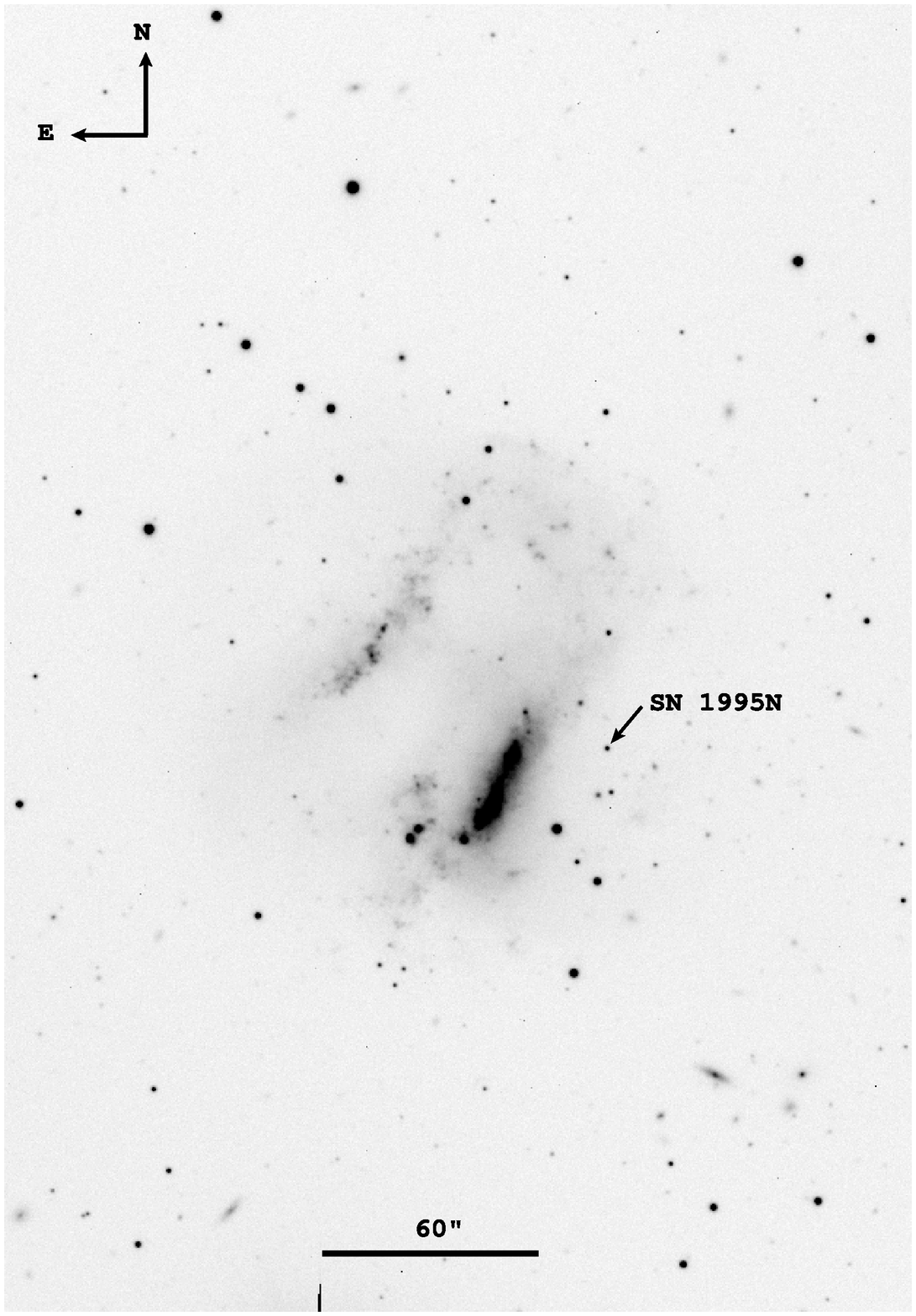}
\caption{VLT $R$-band image of MCG--02-38-017 and SN 1995N from 11 May 1999.}
\label{fig1a}
\end{figure}

\begin{figure}  
\plotone{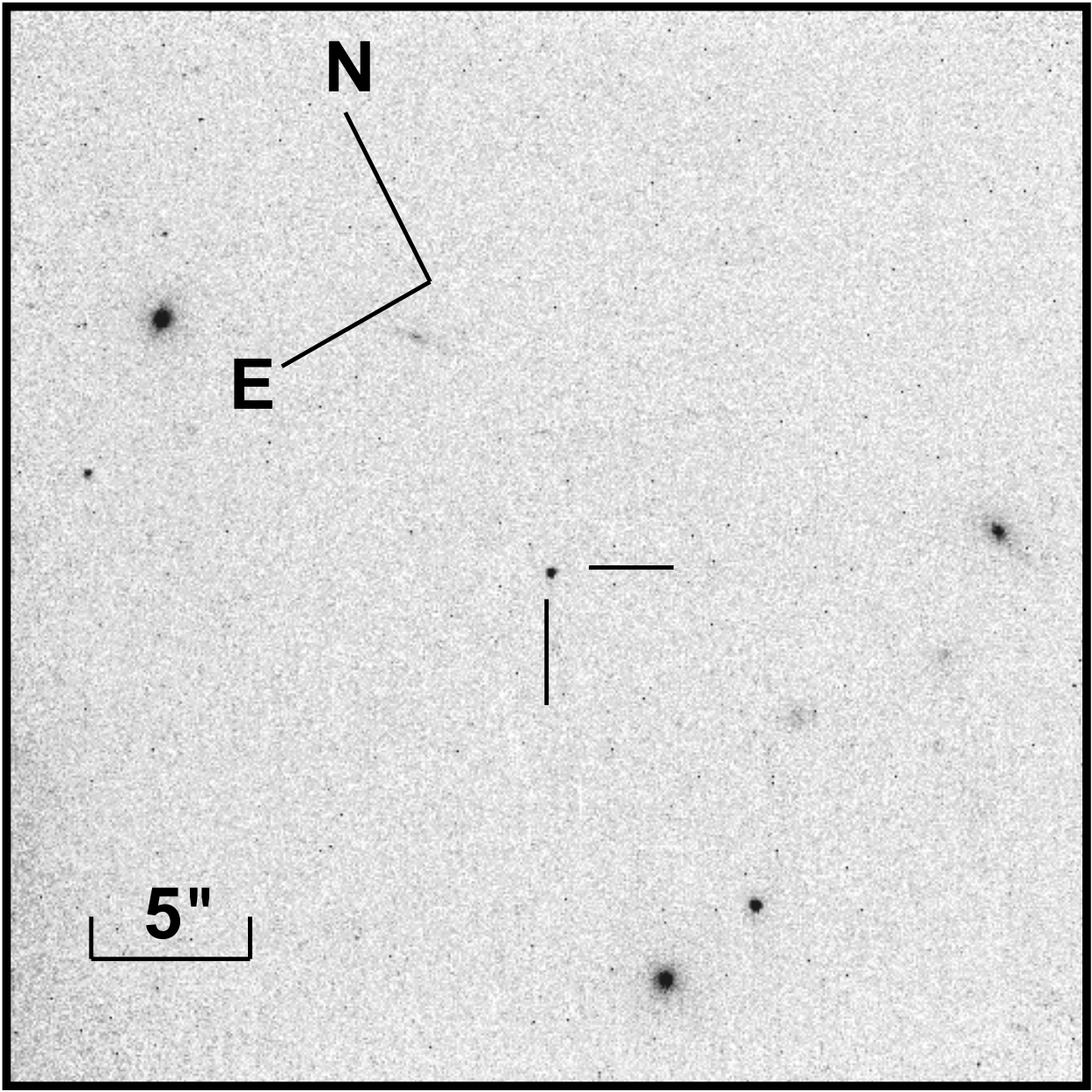}
\caption{{\it HST} PC F814W image of SN 1995N from 22 July 2000. The SN 
has $I \approx 20.5$ mag.}
\label{fig1b}
\end{figure}

  Figure \ref{fig1a} shows a 3-minute $R$-band image of MCG--02-38-107 and SN 1995N
taken with the FORS1 instrument on the ESO Very Large Telescope (VLT) 
on 11 May 1999 (UT dates are used throughout this paper).
Figure \ref{fig1b} is an {\it HST} WFPC2 $I$-band (F814W) image of SN 1995N
obtained on 22 July 2000; the supernova is easily visible at $\sim 20.5$ mag.
These two images demonstrate that the supernova occurred in a region relatively
free of background emission and dust.


The main observations of SN 1995N presented in this paper consist of
ground-based optical spectra obtained on nine epochs and ultraviolet (UV) {\it
HST} spectra on one epoch.  The log of the observations is given in Table
\ref{tab1}, where the wavelength ranges, UT dates, and approximate epoch since
explosion are summarized. The assumed explosion date of 4 July 1994 is
uncertain, but this does not appreciably affect our results because our data
were taken long after the explosion; the spectra cover epochs of 321--1799
days. Standard procedures were followed for the spectral extractions and
reductions, which included sky subtraction, wavelength and flux calibration,
and removal of telluric absorption (e.g., Matheson et al. 2000).

\begin{figure}  
\plotone{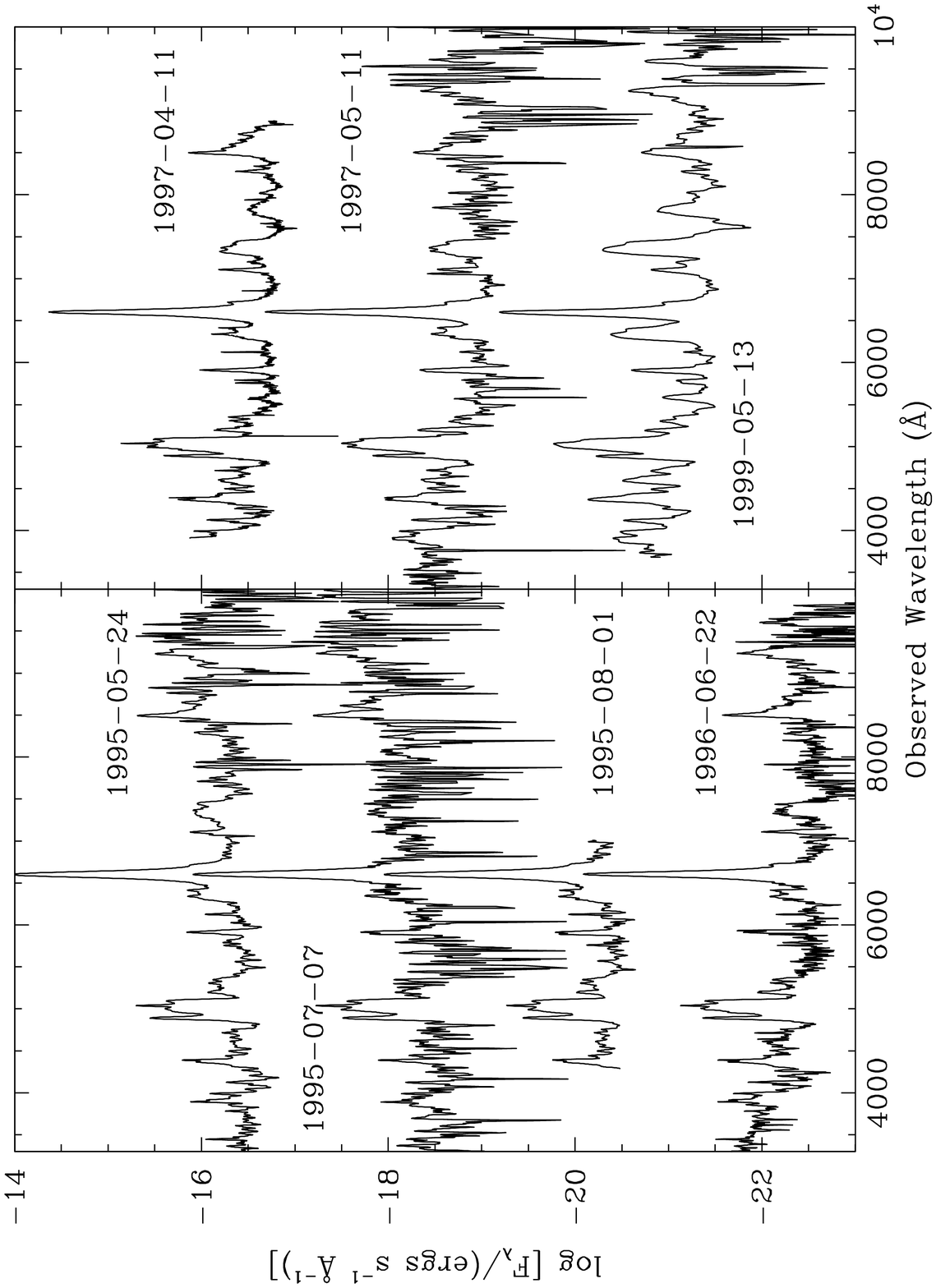}
\caption{Spectra of SN 1995N within 5 years after discovery. All
except the 11 April 1997 Keck spectrum and the 13 May 1999 VLT spectrum were
obtained with the Lick 3-m Shane reflector.  Wavelengths are in the observed
frame. The flux scale is with respect to the upper two spectra, while
all the lower spectra are displaced by two decades with respect to each other.}
\label{fig2}
\end{figure}

  The ground-based observations fall in two main groups: within $\sim 2$ years
after discovery, and $\sim 4$ years after discovery. The former were obtained
primarily with the Kast spectrograph (Miller \& Stone 1993; $2 \arcsec$ slit) on the 3-m Shane
reflector at Lick Observatory, but the Low Resolution Imaging Spectrometer
(LRIS; Oke et al. 1995; $1 \arcsec$ slit) on the Keck-I telescope was used to acquire an
excellent spectrum (day 1007) nearly contemporaneously with one of the Lick
spectra (day 1012). All of the Lick and Keck spectra  (except
day 1012) are
shown in Figure \ref{fig2}. One can see that the spectrum evolved slowly during
the time interval shown. In this paper, we present a detailed analysis of only
two of these spectra (days 716 and 1007, which are among the two best ones).

The very late-time ($\sim 4$ years) ground-based spectra were taken through a $1
\arcsec$ slit with FORS1 on the VLT at ESO. The day 1769
spectrum was obtained with gratings 300V and 300I, giving a resolution of $\sim
13$~\AA.  For the day 1799 spectrum the 600B, 600R, and 600I grisms were used,
giving a higher resolution of $\sim 5$~\AA.  The day 1769 observing was done in
good conditions, while the data from day 1799 were taken through clouds.  The
latter spectrum has higher signal-to-noise ratio ($S/N$) than the former, however, so
emission-line measurements are reported for it (scaled to the flux level of the
day 1769 spectrum).

The {\it HST} spectra on day 943 were obtained with the Faint Object
Spectrograph (FOS) with the G400H, G270H, and G160L gratings. Total
exposures times were 2860, 4770, and 2110~s, and the resolution was
3.0, 2.1, and 6.9~\AA\ per diode, respectively. The spectra were
calibrated by the FOS pipeline at the Space Telescope Science
Institute.

\begin{figure}  
\plotone{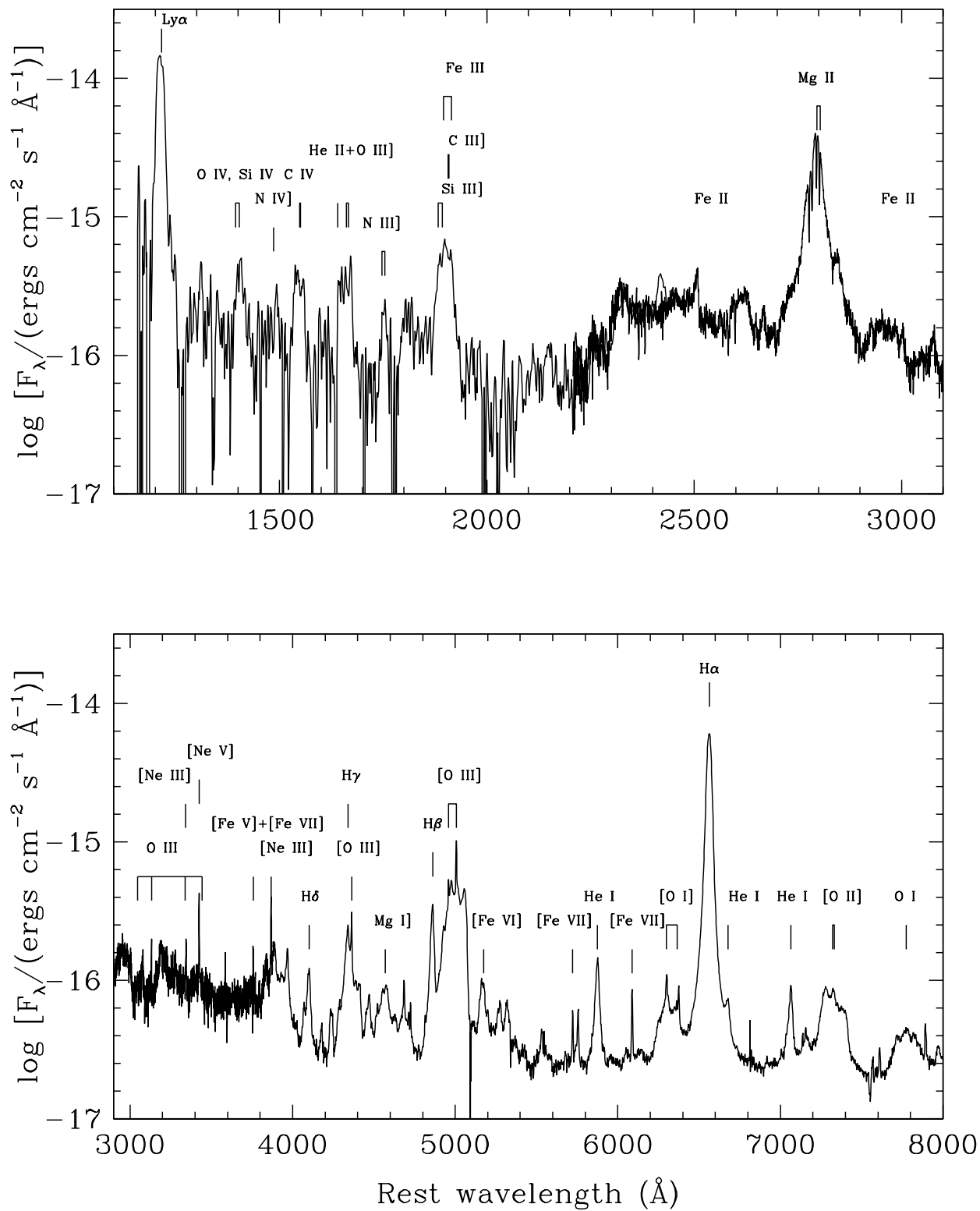}
\caption{Combined day 943 ({\it HST}) and day 1007 (Keck) spectra of SN 1995N, with
identifications of the strongest lines. We have omitted identifications of the many Fe~II
lines in the UV and visible ranges. Wavelengths are in the rest frame
of the supernova.}
\label{fig3}
\end{figure}

All spectra, except the VLT and {\it HST} spectra, were calibrated to the 
$V$-band and $R$-band photometry of Baird \etal (1998). 
The Keck spectrum (day 1007) has the highest $S/N$, and in Figure
\ref{fig3} we show it together with the {\it HST} spectra (day 943). The Keck and {\it
HST} spectra were taken at different epochs, but from the day 1037 Lick
spectrum we find a smooth evolution, and we normalize the flux level of the
overlapping region between the {\it HST} G400H spectrum and the Keck spectrum
to that of the Keck epoch.


The Dirbe/IRAS maps (Schlegel, Finkbeiner, \& Davis 1998) give $E_{B-V} = 0.11$
mag for the Galactic reddening of SN 1995N.  In addition, there could be
internal reddening from both the host galaxy and the circumstellar
environment. The host galaxy is a blue compact galaxy, a class which is
generally dust-poor, although H~II regions can show substantial reddening (Kong
\& Cheng 1999). The fact that the supernova is located in the outskirts of the
galaxy (Figures \ref{fig1a}, \ref{fig1b}) also makes it natural to expect a relatively low
extinction. From the relative strengths of the higher members of the H~I and He~I
recombination series, there is no indication of strong reddening
(\S~\ref{sec_result}).  The normal intensity ratio of UV  O~III] lines
relative to the optical also indicates this. In the following, we use $E_{B-V}
= 0.11$ mag for the total reddening.

\section{KINEMATIC COMPONENTS}
\label{sec_kin}

 Figure \ref{fig3} shows that there are lines spanning a large range of
velocities, and we can distinguish at least three kinematic components in the
supernova. The narrow, unresolved lines are clearly distinct, while the other
two components are less obvious and are shown in the Mg~II $\wll 2796, 2803$,
[O III] $\wll 4959, 5007$, and H$\alpha$ lines at 943--1007 days on the same
velocity scale in Figure \ref{fig4}.  The H$\alpha$ and [O~III] profiles are
from the Keck day 1007 spectrum and the Mg~II profile from the {\it HST} spectrum. The
H$\alpha$ and [O~III] lines have quite different profiles, with the former
having a fairly peaked profile, but with weak wings reaching very high
velocity, $\sim 10,000\kms$, while the latter has a more boxy profile reaching
a velocity of $\sim 5000\kms$. Another strong indication of two separate
high-velocity components is the different temporal evolution of the [O~I], [O~II],
and [O~III] lines compared to the others (\S~\ref{sec_time}). Here we discuss
these three components individually.

\subsection{Narrow Lines}
\label{sec_narrow}

   The narrowest lines in the spectra are unresolved, implying FWHM $\lsim
500 \kms$.  They are listed in Table \ref{tab2}, where it can be seen that the
emission is dominated by ions with moderately high ionization.  Some of the
lines provide useful diagnostics for the emitting region, and we begin by
considering them.

{\bf [O~III]}.  The [O~III] $\wll$4959, 5007 lines are by far the strongest
narrow lines and, together with [O~III] $\lambda$4363, provide a useful
constraint on the electron temperature $T$ and the electron density $n_e$.  The
relatively low ratio of $I(\wl 4959 + \wl 5007)/I(\lambda 4363) = 5.4$ implies
that collisional deexcitation of the $^1D_2$ level is important, so that the
ratio depends on density as well as temperature (e.g., Filippenko \& Halpern
1984; Osterbrock 1989).  The densities corresponding to various assumed
temperatures are in the range $7.3\times10^5 < n_e < 7.5\times 10^7 \cmc$ for
$0.8 < T_4 < 4$ (or $7.3\times10^5 < n_e < 2.0\times 10^6 \cmc$ for $0.8 < T_4
< 2$),
where $T_4=T/(10^4~{\rm K})$.  On day 1799 the VLT spectrum shows a similar
value, $I(\lambda 4959 +\lambda 5007)/I(\lambda 4363) =6.5 $, which gives
$7\times10^5 < n_e < 4\times 10^6 \cmc$ for $1 < T_4 < 2$.

{\bf O~III Bowen lines}.  In the near-UV spectral region, several Bowen
fluorescence lines are seen, most notably O~III $\wll 3047, 3133, 3340, 3444$.
We measure $I(\wl 3047)/I(\wl 3132) \approx 0.30$ and $I(\wl 3444)/I(\wl 3132)
\approx 0.25$. Based on the Bowen process, Kallman \& McCray (1980) give ratios
of 0.18 and 0.28 for these lines, respectively. Considering that the
uncertainties in the fluxes of both lines are at least a factor of two, the
agreement is satisfying. The $\wl 3340$ line is severely blended with [Ne~III]
$\wl 3342$ and [Ne~V] $\wl 3345$, and no reliable flux estimate can be
obtained.

The ratio $I({\rm O~III}~ \wl 3132)/I({\rm He~II}~ \wl 4686)$ can be
used to estimate the Bowen yield for He~II $\wl 304$.  Equation (25)
in Kallman \& McCray (1980), together with $I(\wl 3444)/I(\wl 3132) =
0.28$, gives a Bowen yield of $y_{\rm B} = 0.28 \times I({\rm O~III}~
\wl 3132)/I({\rm He~II}~ \wl 4686)$. Correcting for reddening, we find
$I({\rm O~III}~ \wl 3132)/I({\rm He~II}~ \wl 4686) = 1.05$, and
therefore $y_{\rm B} \approx 0.3$ for SN 1995N at day 1007.  As a
comparison, Schachter, Filippenko, \& Kahn (1990) find a mean $y_{\rm
B} \approx 0.24$ for a sample of Seyfert nuclei, while Kallmann \&
McCray find that planetary nebulae have $y_{\rm B} \approx 0.37$. The
dispersion in both values is considerable.

{\bf [O~II]}.  The spectra show that [O~II] $\lambda\lambda$3726, 3729 are
not detected, but that [O~II] $\lambda\lambda$7320,
7330 are weakly present with narrow components.  The fact that
$I(\lambda 3726 + \lambda 3729)/I(\lambda 7320 + \lambda 7330) <1$
implies that $n_e > 3 \times 10^4 \cmc$, provided that $T < 30,000$ K.

{\bf [N~II]}.  The auroral [N~II] $\lambda$5755 line is present, but the
nebular lines are superposed on the very strong H$\alpha$ emission and it is
not possible to obtain strong limits on them.  We estimate $I(\lambda 6548
+\lambda 6583)/I(\lambda 5755) < 12$, which implies that $n_e > 6 \times 10^4
\cmc$ for $T \approx 20,000$ K (Osterbrock 1989).

{\bf [Fe~VII]}.  The observed lines of [Fe~VII] are $\lambda\lambda$3586, 3759,
5159, 5276, 5721, 6087.  As discussed by Nussbaumer \& Storey (1982) and Keenan
\& Norrington (1987, 1991), several of the [Fe~VII] lines provide diagnostic
information on the emitting region.  There is temperature sensitivity of the
lines, and collisional deexcitation causes transitions from the low-density
limits to the high-density limits in the range $n_e \approx 10^6-10^8 \cmc$.

The observed, reddening-corrected ratio $I(\lambda 3759)/I(\lambda 6087)
\approx 0.83$ in the Keck spectrum (day 1007), and $\sim 1.1$ in the 12 June
1999 VLT spectrum (day 1799).  From Figure 2 of Keenan \& Norrington (1987), this sets a
limit $T > 17,000$ K for any density on day 1007, and $T > 22,000$ K for day
1799.

In the day 1799 spectrum, with the highest resolution, there are three
distinct peaks at the positions of [Fe~VI] $\wl 5146$, [Fe~VII] $\wl 5159$, and
[Fe~VI] $\wl 5176$. Because of blending of these lines, it is difficult to
obtain accurate fluxes, and depending on where we estimate the base of the line
we get a ratio in the range $0.20~ \lsim ~I(\lambda 5159)/I(\lambda 6087) ~\lsim
~0.73$ and $0.25~ \lsim ~I(\lambda 5159)/I(\lambda 6087)~ \lsim ~1.1$ on days 1007
and 1799, respectively.  From Figure 1 of Keenan \& Norrington (1987), the
lower limits imply that $n_e~ \lsim ~3 \times 10^6 \cmc$ if $T < 60,000$ K.  The
lines at $\lambda$4942 and $\lambda$4989 are contaminated by the strong complex
of [O~III] $\lambda\lambda$4959, 5007, with the result that interesting
constraints on $T$ and $n_e$ cannot be obtained from these lines (Keenan \&
Norrington 1991).

Overall, the above results point to a relatively high density for the
narrow-line gas, which suggests a circumstellar rather than interstellar
origin.  If the [O~III] and [Fe~VII] lines are assumed to come from gas with
similar physical properties, the implied characteristics are $n_e \approx 2
\times 10^6 \cmc$ and $T \approx 20,000$ K, although there is some evidence
that the [Fe~VII] emission comes from a somewhat hotter region. It is likely,
because of the low velocity and the relatively low pressure, that this gas
is the unshocked circumstellar wind. Presumably it was photoionized by the
UV and X-ray radiation emitted by the supernova right after shock breakout.

One can derive a lower limit to the radial distance of this gas by taking the
product of the highest observed supernova ejecta velocity and the estimated
age.  As discussed below, the highest velocity is $\sim 10^4 \kms$. The wind
electron density in front of the shock then becomes
\begin{equation}
n_e = 5.0 \times 10^4\left(\dot M\over 10^{-4}\ml\right)
\left( v_w\over 10 \kms\right)^{-1}
\left(t\over 1000 {\rm~days}\right)^{-2} \cmc,
\end{equation}
where $\dot M$ is the mass-loss rate from the progenitor star, $v_w$
is the wind velocity, and it has been assumed that H and He are fully
ionized.  The reference value of $\dot M$ chosen here is close to the
highest values that have been observed for Galactic stars and to the
strong radio emitter SN 1979C (e.g., Lundqvist \& Fransson 1988).
Either the progenitor star of SN 1995N was losing mass at an
extraordinary rate, or the density of the observed gas is higher than
the average density in the circumstellar wind.

The gas density is sufficiently high that the recombination and excitation time
scales are less than the supernova age, and approximate equilibrium is expected.
The ionizing source is the X-ray emission, which had an unabsorbed X-ray
luminosity of $\sim 1 \times 10^{41}\ergs$ in the 0.1--10 keV band during July
1996 -- January 1998 (Fox et al. 2000).  The {\it ASCA} observations give a
temperature estimate of 9~keV if the emission is thermal (Fox et al.  2000).
Models of X-ray nebulae have been presented by Tarter, Tucker, \& Salpeter
(1969) and Kallman \& McCray (1982).  The latter models assume a constant
density surrounding a central source and generally take $T=10$ keV for the
photoionizing source, but the calculations can be used to approximate the
current situation, especially because the [Fe~VII] lines are likely to be
formed in a region optically thin to the ionizing continuum.  The ionization
properties depend on the ionization parameter, which can be defined as $\xi =
L/nr^2$, where $L$ is the luminosity of the ionizing source.  Taking $L=1\times
10^{41}\ergs$, $n=2\times 10^6\cmc$, and $r=8.6\times 10^{16}$ cm, we find $\log
\xi =0.8$, which is very close to the ionization parameter where Fe~VII is
expected, for a $T=10$ keV source (Kallman \& McCray 1982). A lower temperature
for the continuum source would increase the estimated $\xi$ for a fixed $L$.
The gas temperature of this region in the photoionization models is $\sim
15,000-20,000$ K, consistent with the observations.  This comparison provides
support for the overall model and, in particular, for the estimated radial
distance of the emitting region.

The intensity of the narrow lines gives another constraint on the physical
conditions.  The hydrogen lines are the best to use because there are no
abundance uncertainties.  While no narrow H~I lines can be seen in the day
1007 Keck spectrum because of the strong broad component, the higher resolution
day 1799 VLT spectrum clearly shows H$\alpha$, H$\beta$, and H$\gamma$. For
H$\beta$, we find a luminosity of $5.3 \times 10^{36} \ergs$ at 1799 days.  If
the emitting region is assumed to be at radius $r$ with a thickness $\Delta r$
and a covering factor $f$, we have
\begin{equation}
f{\Delta r\over r}\approx 1.6\times10^{-3} \left(r\over 1.3\times
10^{17}{\rm~cm}\right)^{-3} \left(n_e\over
10^6\cmc\right)^{-2},
\end{equation}
where the emissivity of H$\beta$ has been taken for $T=20,000$ K (Osterbrock
1989). Although the density is uncertain, it is clear that the observations
imply a small filling factor.

Another indication of the small filling factor of the $n_e = 2 \times 10^6
\cmc$ gas is the detection of a substantial X-ray flux by {\it ROSAT}.  If the
gas were pervasive at $r = 8.6 \times 10^{16}$ cm, the resulting hydrogen column
density, $N_H\approx 10^{23} \cms$, would give rise to the complete absorption
of the soft X-ray flux.  The gas is not fully ionized and thus can effectively
absorb the X-ray emission.  In their estimate of the X-ray luminosity, Fox et
al. (2000) take $N_H = 0.8 \times 10^{21} \cms$ based on the expected Galactic
absorption.

Most of the other ions that are observed in SN 1995N require an ionization
parameter that is similar to that of Fe~VII.  The ions Fe~VI through Fe~XI,
which are observed, occur over a relatively small range of ionization parameter
(e.g., Kallman \& McCray 1982).  However, the presence of narrow [O~I] emission
implies the existence of a separate, lower-ionization component.  The region
could be of low ionization because of an especially high density, but it is
more likely that the region is neutral because the ionizing photons have been
used up in the H~II region.

\subsection{Intermediate Component}
\label{sec_interm}

The intermediate-width component is best exemplified by the [O~III]
$\lambda\lambda$4959, 5007 (Figure \ref{fig4}), [O~II]
$\lambda\lambda$7320, 7330, and [O~I] $\wll 6300$, 6363 emission
lines, which all show box-shaped profiles extending to $\sim 5,000
\kms$. The strongest feature in this component at all epochs is the
[O~III] $\lambda\lambda$4959, 5007 blend. Other likely members of this
component are the high-ionization far-UV lines.  Mg~I] $\lambda$4571 is
a more uncertain member, showing a red deficit if it is correctly
identified; alternatively, it could be a blend of lines, in particular
[Fe~II] $\wll 4556$, 4583. The strongest argument for including Mg~I]
in this component is its temporal evolution, which is similar to that
of the [O~I] and [O~III] lines (\S~\ref{sec_time}). There is no
indication of hydrogen or helium with this type of profile.

  Table \ref{tab3} lists the fluxes of the lines that we designate as
having line profiles coming from the intermediate component.  As with
the narrow lines, the intensity ratios of some of these lines provide
constraints on the electron density and temperature.

{\bf O~III}.  The [O~III] $\lambda$4959 and $\lambda$5007 lines are strongly
blended with each other, and the $\lambda$4959 line with H$\beta$.  A feature
at $-1,800 \kms$ appears in both the $\lambda$5007 and $\lambda$4959 lines, and
the ratio of line strengths appears to be consistent with the 3:1 ratio
expected from their spontaneous emission probabilities.  The red edge of the
$\lambda$5007 line extends to $5,000 \kms$, but not further.  The $\lambda$5007
line also has a strong feature to the red at $+3,000 \kms$; however, it does
not appear in [O~III] $\lambda$4363 or in other lines of this kinematic
component, so there is a question of whether the feature originates from
[O~III] or is a blend. Its rapid drop in the late-time spectra (see
\S~\ref{sec_time}) argues for the blend interpretation.  Likely candidates for
this feature are He~I $\wl 5049$ or [Fe~II].

Broad [O~III] $\lambda$4363 emission is present, although the blue
side of the line is disturbed by H$\gamma$ emission.  Neglecting the red
feature in the $\lambda$5007 line, we find $I(\lambda 5007 + \lambda
4959)/I(\lambda 4363) \approx 8.2$ from the day 1007 Keck spectrum.
In the far-UV, the O~III] $\wll 1660, 1666$ intercombination lines are
among the strongest UV lines. Based on the wavelength, it is unlikely
that the He~II $\wl 1640$ line contributes much to this feature. We find a
reddening-corrected ratio [O~III] $I(\wll 4959, 5007)/ I(\wll 1660,
1666) \approx 4.0$.


{\bf O~II}.  The [O~II] $\lambda\lambda$7320, 7330 doublet is less affected
by blending and yields a clearer line profile.  The feature at $-1,800 \kms$ is
present, but that at $+3,000 \kms$ is not.  The line profile is asymmetric,
although the maximum velocity is $4,500 \kms$ to both the blue and the red. As
we show in \S~\ref{sec_result}, the asymmetry is mainly caused by blending with He~I
$\wl 7281$.

{\bf O~I}. The [O~I] $\wll 6300$, 6364 doublet is severely affected by the
blue wing of H$\alpha$. Narrow components of the [O~I] doublet, as well as
[Fe~X] $\wl 6374.5$, also complicate the profile. In \S~\ref{sec_synt} we show
that the line profile is compatible with that of the other oxygen lines. The
auroral [O~I] $\wl 5577$ line is likely to be present, although it may be
distorted by the corresponding night-sky line. We therefore only consider its
flux as an upper limit.

The presence of the O~I recombination line at $\lambda$7774 is of
special interest because it should be accompanied by strong H~I
recombination emission with the same line profile, if the emission is
from material with ``cosmic'' abundances, since hydrogen is more
abundant than oxygen by a factor of $10^3$ in such material.  The
absence of the corresponding hydrogen emission suggests that the
emitting gas is oxygen rich.  The O~I recombination line at
$\lambda$8446, which is expected to have a strength comparable to
the $\lambda$7774 line, is blended with Fe II lines, and is also
affected by fluorescence (\S~\ref{sec_fluor}).  The emission may be from the
oxygen-rich ejecta in the supernova and may extend to relatively high
velocity because of hydrodynamic instabilities with the overlying
layer (Fryxell et al. 1991) and/or a high expansion velocity of the
oxygen core (\S~\ref{sec_nature}).

The lines observed in the short-wavelength G160L {\it HST} spectrum are noisy,
so there is uncertainty in the line identifications.  The line widths, $ \pm
(3500 - 4000) \kms$, point to memberships of most of the UV emissions with the
intermediate kinematic component. The line profiles of the strongest lines
(C~IV, O~III, and Si~III/C~III) are all consistent with flat-topped profiles of
width $\pm 3800 \kms$. Except for the $\wl 1900$ feature, which has a central
peak, the lines appear symmetric within the noise level. The C~IV and Si~IV
lines are also affected by interstellar absorption in the Galaxy and the host
galaxy.

By far the strongest line is Ly$\alpha$, whose profile is distorted by
geocoronal Ly$\alpha$, and possibly by N~V $\wl 1240$.  The C~IV $\wll 1548$,
1551 doublet is clearly seen with a boxy profile.  The same is true for the
feature at $\lambda$1900, which can be identified as a blend of Si~III]
$\lambda$1892 and C~III] $\lambda\lambda$1907, 1909. The N~III] 
$\wll$1744--1754 lines are clearly present, although their flux is uncertain. The
N~IV] $\wl 1486$ line is near the limit of detection.

In Figure \ref{fig4a} we show the density and temperature constraints from all
the oxygen-line ratios of the {\it HST} and Keck day 1007 spectra together. We have
used a reddening correction $E_{B-V} = 0.11$ mag to obtain $I({\rm [O~I]}~ \wl
7774)/I({\rm [O~II]}~ \wll 7320, 7330) = 0.43$, [O~III] $I(\wll 4959,
5007)/I(\wl 4363) = 8.2$, and [O~III] $I(\wll 4959, 5007)/I(\wl 1664) =
4.0$. The [O~I] $I(\wl 5577)/I(\wll 6300, 6364)$ ratio is only an upper limit,
$\lsim 0.1$, and the curve shown in Figure \ref{fig4a} therefore only marks an
upper boundary to the temperature. Implicit in this analysis is that the O~I
and O~III zones are of similar extent, temperature, and electron density.
While the lines coming from the O~II and O~III zones are all compatible with $T
\approx 7000$ K and $n_e \approx 3 \times 10^8 \cmc$, the [O~I] lines have to
arise in a lower temperature and/or lower electron density zone. As shown in
Chevalier \& Fransson (1994, hereinafter CF94), both these alternatives are
quite natural, because the [O~I] lines are expected to arise in a partially
neutral zone of lower temperature (see, e.g., CF94 Figures 4 and 5).

The velocities observed in the intermediate component suggest that this emission is from
freely expanding ejecta.  The ejecta are photoionized by X-rays from the
circumstellar interaction.  This process has been modeled by CF94, and there is
general agreement between the lines that are predicted to be brightest and
those that are observed here.


\subsection{Broad Component}
\label{sec_broad}

\begin{figure}  
\plotone{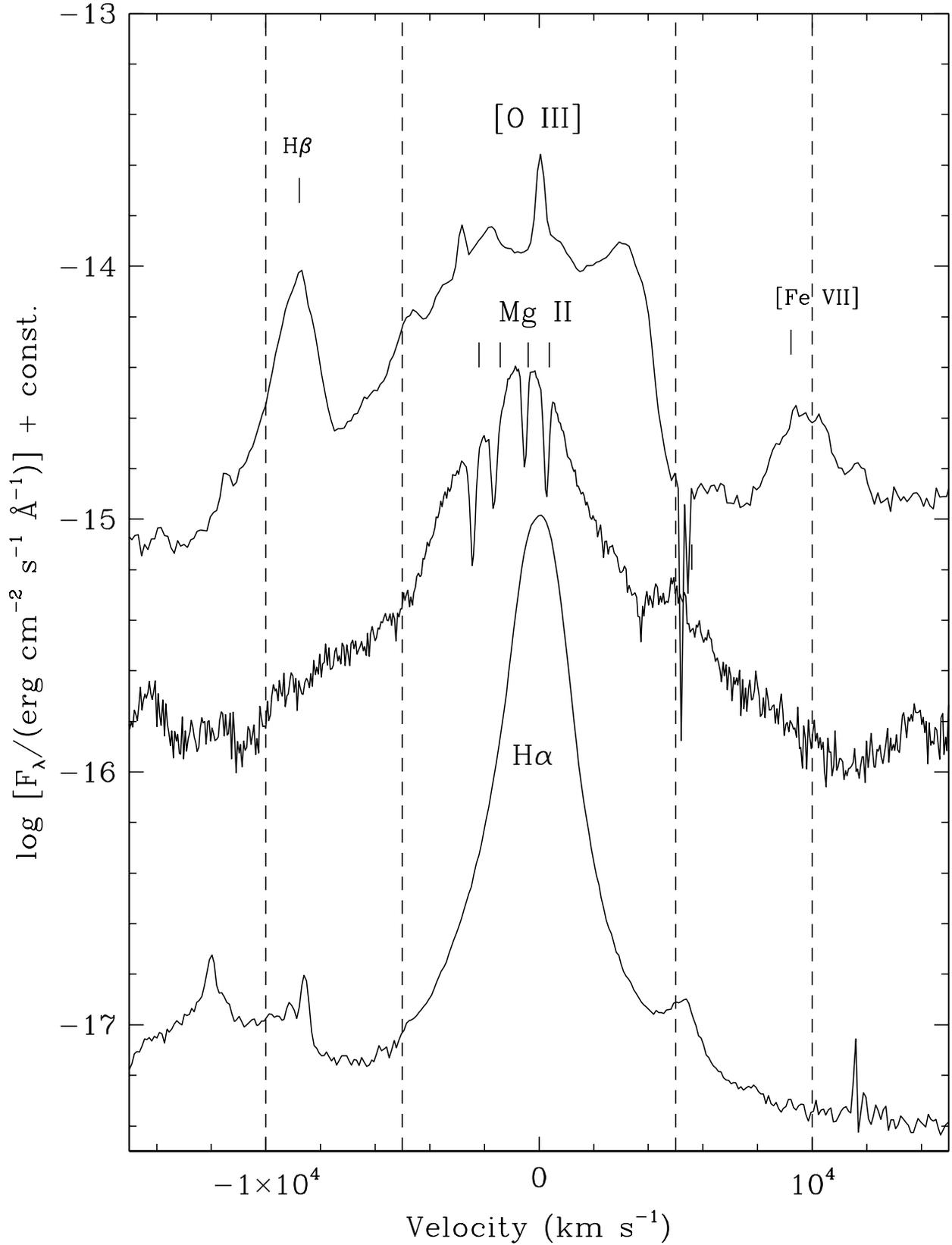}
\caption{Line profiles of Mg~II $\wll$2796, 2803, [O~III] $\wll$ 4959,
5007, and H$\alpha$.}
\label{fig4}
\end{figure}

\begin{figure}  
\plotone{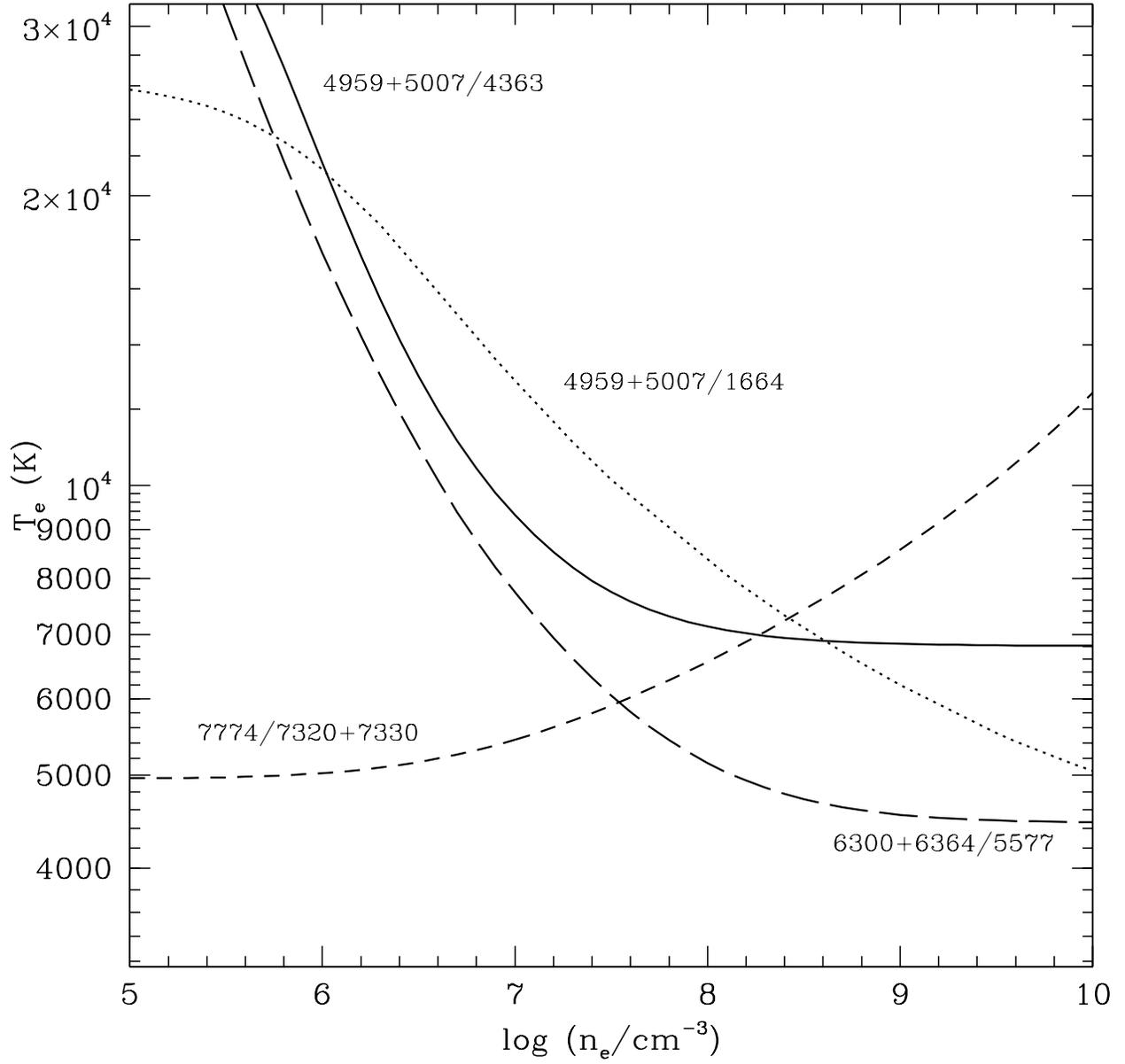}
\caption{Temperature vs. density plot for the oxygen lines. The [O~I]
$\wll$6300, 6364/$\wl$5577 curve marks only an upper limit to the
temperature for a given electron density.}
\label{fig4a}
\end{figure}

Because of the strong H$\alpha$ line, the broad component dominates the optical
emission from the supernova. The total luminosity in H$\alpha$ was $\sim 2.3
\times 10^{40} \ergs$ on day 1007, using $E_{B-V} = 0.11$ mag. The line profile
of H$\alpha$ is shown in Figure \ref {fig4}.  There is no obvious distinction
of different components in H$\alpha$, as we demonstrate in \S~\ref{sec_result},
and no clear evidence for any velocity structure within the line, although a
slight emission deficit can be seen on the red side of the line.  The FWHM of
the line is $\sim 1600 \kms$ on day 1007. The wings are very extended and
reach at least 10,000 $\kms$.


Table \ref{tab4} gives a list of the lines that have profiles similar to that
of H$\alpha$, although we can trace the high-velocity wings above $\sim 3000
\kms$ only for H$\alpha$.  Besides H$\alpha$, the best-defined line profile
from this component is He~I $\wl 5876$, which is very similar to H$\alpha$ up
to $\sim 3000 \kms$, where the line is lost in the noise. The broad
component is dominated by lines of H~I, He~I, Mg~II, and Fe~II.

The interpretation of the Mg~II $\lambda\lambda$2796, 2803 doublet is complicated
by absorption features arising in gas in our Galaxy and in the host
galaxy of SN 1995N.  The line peak is at negative velocities, at $\sim 2791$ \AA,
which may be a result of resonance-line scattering of the $\wl 2796$ photons by
the $\wl 2803$ component, as well as the influence of the interstellar
absorption lines. The line appears considerably broader than H$\alpha$, and
shows a clear deficit of emission to the red of line center (Fig. \ref{fig4}).
As discussed in \S~\ref{sec_fluor}, the extended profile is likely to be caused
by blending with Mg~I $\wl 2852$ and with the strong Fe~II resonance lines in
this region of the spectrum, expected to be present  on either
side of the wings (multiplets 399, 391, 380 at $\sim 2850$~\AA, and multiplet
373 at $\sim 2770$~\AA), due to the Ly$\alpha$ fluorescence.

Although almost all of the many Fe~II lines present are blended with each
other, it appears that they belong to the broad kinematic component. The only
reasonably isolated Fe~II line is $\wl 7155$, whose profile is well fitted by a
scaled H$\alpha$ profile. In the optical spectra, the region 4000--5500~\AA\
contains a large number of Fe~II lines, often seen in supernovae and active
galactic nuclei (AGNs; e.g., Filippenko 1989; Osterbrock 1989).  Comparing with
spectra of narrow-line Seyfert 1 galaxies such as I~Zw~1 (Phillips 1976, 1977),
and the symbiotic star RR Tel (Crawford \etal 1999), we have identified members
of Fe~II multiplets 14, 27, 28, 27, 42, 48, 49, and 74 (Table \ref{tab4}). The
lines listed in Table \ref{tab4} are those that can be identified with
reasonable certainty.  Most of the other members of the multiplets are blended
with strong lines. One should also note that the fluxes of all Fe~II lines,
with the possible exception of the $\wl 7155$ line, are highly uncertain.

In the UV, the plateau between $\sim 2300$~\AA\ and $\sim 2800$~\AA, except for
Mg~II, is dominated by Fe~II resonance lines. Because of the extreme blending
of lines, which creates a pseudo-continuum, we do not attempt any complete
identification in this spectral region. We simply note that there are some
distinct features such as Fe~II $\wll 2614-2631$ and Fe~II $\wll 2506,
2508$. The latter lines are of special interest, because they arise from
cascades from highly excited levels, and are usually considered an indicator of
fluorescence. Except for this, the UV part of the spectrum is similar to that
of I~Zw~1 (e.g., Fig. 2d in Laor \etal 1997).

In the near-IR, it is especially interesting to note that many of the features
can only be identified with Fe~II emission based on models of Ly$\alpha$
fluorescence of the transitions (Johansson \& Jordan 1984; Penston 1987; Sigut
\& Pradhan 1998a). Sigut \& Pradhan (1998b), who discuss this process in detail
for AGNs, provide a detailed list of the predicted lines, yielding many
identifications for SN 1995N.  The best example of Ly$\alpha$ fluorescence is
probably the feature at $\sim 9150$~\AA, which is difficult to explain with
other mechanisms. Based on other Paschen lines, a significant contribution of
H~I is unlikely, and He~I lines are also excluded by similar arguments, as well
as by their wavelengths. Another candidate, Mg~II $\wll 9218, 9244$, also has
the wrong wavelengths. 
In \S~\ref{sec_fluor} we
show explicitly how the inclusion of these lines improves the spectral fit.

Finally, we point out the very strong He~I $\wl 10830$ line in the VLT day
1769 spectrum, which is the only spectrum covering this region.

Although there are many lines observed with the high-velocity component, the
only straightforward diagnostic of the physical conditions is given by the lack
of forbidden-line emission.  In the case of the broad-line regions of AGNs,
this is taken to indicate a density $\gsim 10^9 \cmc$ (e.g., Osterbrock
1989). This should be taken with some caution, however, because the gas is only
partially ionized by the hard X-rays (CF94); high-ionization lines like [O~III]
$\wll 5007, 4959$
and C~III] $\wl 1909$ are therefore unlikely to be present. The only forbidden
lines expected are [O~I] $\wll 6300, 6364$, which are suppressed relative to
permitted lines for $n_e~ \gsim
~10^7 \cmc$. As we discuss in \S~\ref{sec_result}, the large H$\alpha$/H$\beta$
ratio, as well as the Ca~II line ratios, strongly argue that $n_e~ \gsim ~10^9
\cmc$, which is a likely lower limit for the broad component.

The emission is presumably from a radius that is similar to the radius of
the narrow-line emission, and the gas is thus exposed to a radiation field that
is comparable to that of the narrow-line gas.  The high density results in an
ionization parameter that is a factor $\gsim 10^3$ smaller than that for the
narrow-line gas, or $\log \xi \la -2$.  The result is low ionization of the
gas, as observed.  The ionization is much less than that in the broad-line
regions of AGNs, where lines like C~IV $\lambda$1549 are observed.


\section{SYNTHETIC SPECTRA}
\label{sec_synt}

 To determine abundances as well as deblend the most important lines of the
intermediate and broad components, we have calculated synthetic spectra based
on the combined UV-optical {\it HST}/Keck spectrum from \S~\ref{sec_obs}.  The
model we employ is a simple two-component model, with constant physical
parameters and abundances in each of the components. The temperature,
densities, and abundances are taken as input parameters. For each of the two
components we assume a given emissivity profile with velocity.  The
emissivities of the H and He lines are given by Case-B recombination, except
for H$\alpha$, which is multiplied by some factor to account for collisions and
optical-depth effects. 
The He~I $\wll$3889, 7065 lines are also adjusted by a common
factor to account for optical-depth effects (see Osterbrock 1989, p. 105). All
lines from metals are calculated as multilevel atoms for a given temperature
and density. The ionization fractions for a given ionization stage (such as C
III, N III, O III, and Si III) are assumed to be the same. The continuum is due
to H I, He I, and He II bound-free and two-photon continua, consistent with the
recombination lines.  We do not attempt to include the many Fe II lines
present, nor lines from the narrow component.
The temperature and ionization are not calculated self-consistently,
as in CF94, so only the relative line fluxes for a given ion, or group of ions,
are testing the model. The model should therefore mainly be viewed as a
diagnostic tool.

For the line profiles we adjust the variation of the emissivity with velocity
so that a best fit is obtained (eq. \ref{eq_a}). For spherical symmetry and for
a velocity $V(r) = V_{\rm max}(r/R_{\rm max})$, the flux at wavelength
$\lambda$ is given by
\begin{equation}
F_{\nu}(\lambda) = 2 \pi \int^R_x S(r) (1 -\exp[-\tau(r)]) r \, dr,
\label{eq_a}
\end{equation}
where $S(r)$ is the source function and $\tau(r)$ is the optical depth. In the
optically thin limit $S(r) [1 -\exp(-\tau)] \approx j(r)$, where $j$ is the
emissivity. The limits of integration are given by $x = c t (\lambda -
\lambda_0)/\lambda_0$, and $R = V_{\rm max} t$, where $V_{\rm max}$ is the
maximum ejecta velocity and $c$ is the speed of light.  This characterization
of the line profile is appropriate for freely expanding ejecta; free expansion
is assumed in our synthetic spectral models.
As discussed in \S~6.1.3, this assumption may not apply to the broad
component.
In this case, the line profile should be viewed as an empirical fit
that is applied to the set of lines.

\subsection{Results}
\label{sec_result}

We find that the profile of the H I, He I, Mg II, and Fe II lines can be
well reproduced with an emissivity [or, if $\tau \gg 1$, a source function
$S(r)$] which is constant inside 1000 $\kms$ and decreasing like $j(V) \propto
V^{-4.6}$ outside, out to about $15,000 \kms$.  For the metal lines, except for
Mg~II and Fe II, we have used a constant-emissivity shell with $j =$ constant
for $2500 < V < 5000 \kms$.

\begin{figure}  
\plotone{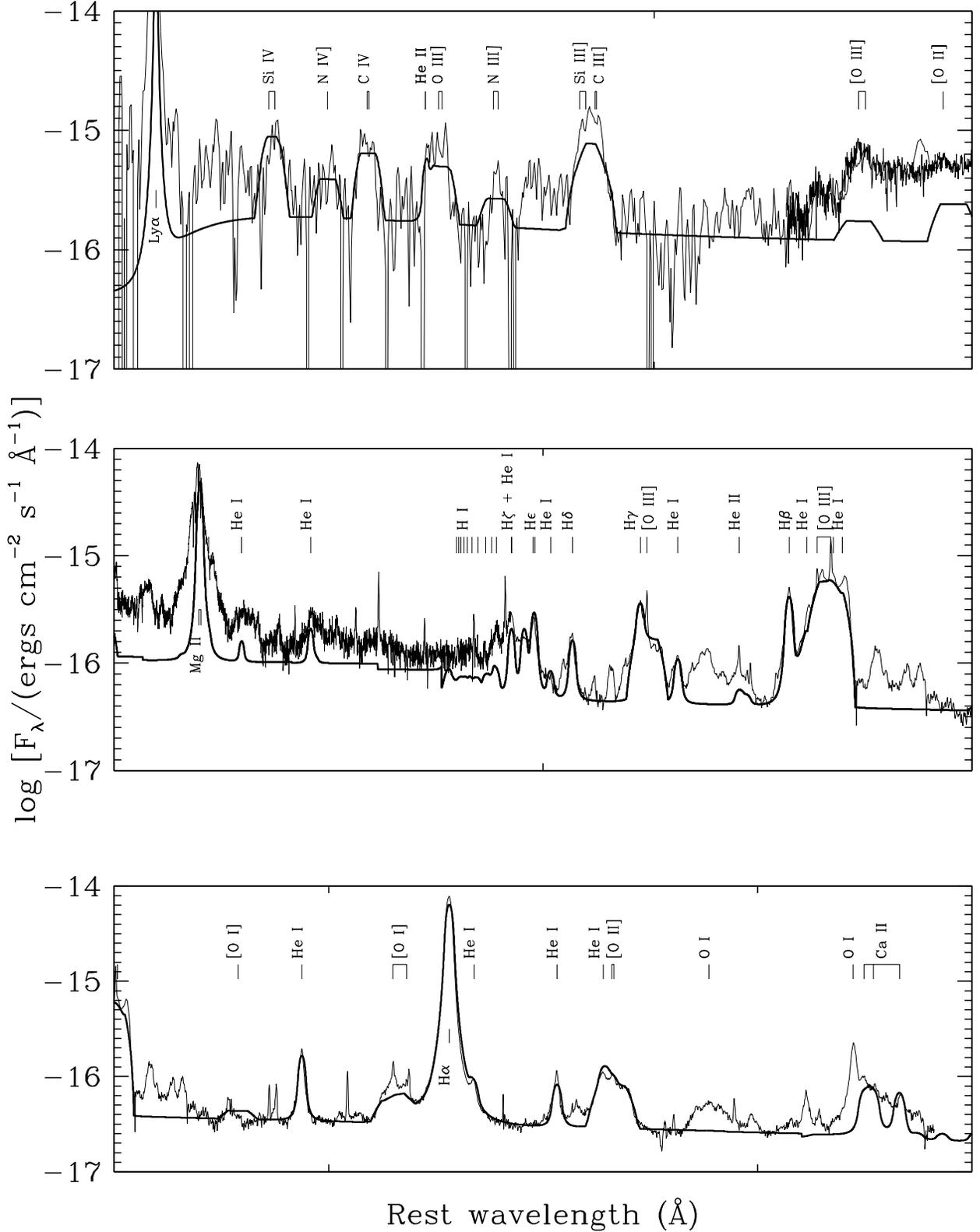}
\caption{Synthetic spectrum superimposed on the day 1007--1037
spectra, corrected for reddening. The model includes H, He, C, N, O,
Mg, Si, and Ca, but no Fe~II lines.  The metallicity is close to solar,
$T = 14000$ K, and $n_e = 4 \times 10^6 \cmc$. Note the severe
underproduction of O~I $\wl$7774, caused by the normal metallic
abundances.}
\label{fig5}
\end{figure}

In Figure \ref {fig5} we show the spectrum of a model where all elements,
including H, He, and metals, have the same density and temperature, $4\times
10^6 \cmc$ and $14,000$ K. The total metallicity is close to solar. Note that
the temperature, density, and metallicity are only indicative, and depend on
the assumed total emissivity, $\int n_e n r^2 dr$. The fit gives a good general
representation of most lines, in terms of both the relative fluxes and line
profiles. In particular, many of the profiles that seem fairly complicated at
first sight are well reproduced as a result of blends by several lines. Good
examples are the [O~I] + H$\alpha$ complex, and the [O~III] + H$\beta$ feature
at 4800--5000 \AA.  The model, however, severely underproduces the strength of
the important O~I $\wl$7774 recombination line.

This could have been anticipated already from Figure \ref{fig4a}, which shows
that the relative [O III] ratios can be reproduced either with a
high-temperature, low-density model, as in Figure \ref {fig5}, or with a low
temperature and a high density. The [O II] $\wll$7320 + 7330/O I $\wl$7774
ratio can, however, only be reproduced with the latter temperature/density
combination. The relative fluxes of the oxygen and hydrogen/helium lines then
require a very high oxygen abundance, unless the O II and O III zones are
unreasonably large compared to the zone responsible for the H I and He I
recombination lines. For the same extent, the relative flux in O I $\wl$7774
and H I lines is given by [$j_{\rm eff}({\rm O~I}~ \wl$7774)$n({\rm O~II})]/
[j_{\rm eff}({\rm H~I}) n({\rm H~II})]$. Because $j_{\rm eff}({\rm O~I}~ \wl$
7774) $\approx 2 \times j_{\rm eff}({\rm H}\beta$), this shows that $n({\rm
O~II})$ and $n({\rm H~II})$ must be of the same order of magnitude.

\begin{figure}  
\plotone{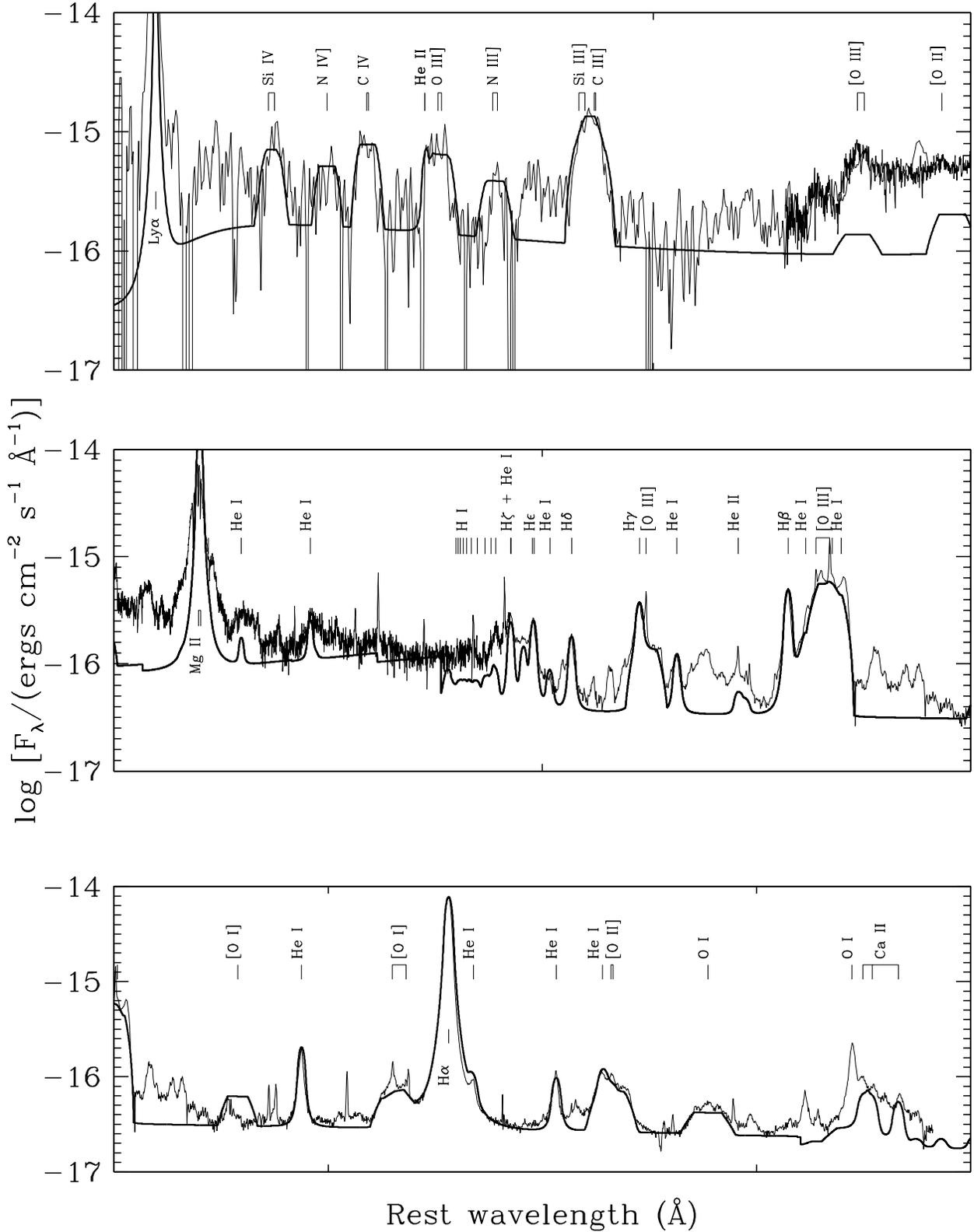}
\caption{Same as Fig. \ref{fig3}, but now with $T = 7500$ K, $n_e =
3 \times 10^8 \cmc$, and a high metal abundance. The O~I
$\wl$7774 line is now well-reproduced relative to the other O~I and O~II lines.}
\label{fig5b}
\end{figure}

As an alternative, we show  a model with a greatly
increased metal abundance (Fig. \ref{fig5b}).  Although these lines are likely to arise in
different ionization zones, we find that we can get an acceptable fit to all
lines with a density $\sim 3\times 10^8 \cmc$ and a temperature $\sim 7,000$ K.
All metal lines, except Mg~II and Fe~II, are assumed to have a shell-like
emissivity profile.  Compared to Figure \ref {fig5} we see that, because of the
high oxygen abundance, now both [O~II] $\wll$7320, 7330 and O~I $\wl$7774, as
well as the strengths relative to the H~I lines, are well reproduced. In
addition, the O~I $\wl$8446 recombination line contributes appreciably to the
feature at $\sim 8450$ \AA. We also note the good fit to the O~I $\wl$7774 line
profile, further strengthening this identification.

To reproduce the observed H$\alpha$/H$\beta$ ratio, the H$\alpha$ strength has
to be increased by a factor of $\sim 8$ above recombination.  The
reddening-corrected Ly$\alpha$/H$\alpha$ ratio is $\sim 2.3$, while the Case-B
ratio is $\sim 24$. The flux of Ly$\alpha$ is, however, highly uncertain,
because the line profile is disturbed by geocoronal Ly$\alpha$, in spite of the
recession velocity of SN 1995N. As the self-consistent models in CF94 show,
this occurs because of collisional excitation in combination with a high
optical depth for H$\alpha$ in the partially ionized zone. As both the
calculations in CF94 and similar models for AGNs (e.g., Kwan \& Krolik 1981)
show, this requires the presence of hard X-rays from the shocked region in
combination with a density $\gsim 10^8 \cmc$.  The higher-order Balmer lines
are well explained by Case-B recombination.

The helium lines are well reproduced by pure recombination, except for He~I
$\wl$3889 and He~I $\wl$7065. The former is down by a factor of $\sim 0.5$, and
consequently the He~I $\wl$7065 line is increased by a factor of $2.0 \times
7065/3889 \approx 3.6$, which strongly indicates that the He~I $\wl$3889 line
is optically thick (e.g., Osterbrock 1989, p. 105).

The relative ratio of the He~I and H~I fluxes results in a He~II/H~II abundance
of $\sim 0.35$ by number. Because the He~II and H~II ionization zones are
likely to coincide (e.g., CF94), the He~II/H~II ratio is probably close to the
elemental He/H abundance ratio.  Strong He~I lines are also present in SN 1986J
(Rupen et al. 1987; Leibundgut et al. 1991) and at early times in many SNe~IIn
(e.g., Filippenko 1997).

The jump at $\wl \approx 3700$~\AA\ is mainly caused by high-order Balmer lines
and the Balmer continuum.  The two-photon continuum dominates in most of the
UV, as well as in the optical. The form of the Balmer continuum is sensitive to
the temperature in the zone where this is formed, and a somewhat better fit is
obtained for $T \approx 9000$ K. In a one-zone model, however, this temperature
gives a worse fit to the metal lines. More detailed models (CF94) show that
both temperature and ionization vary within the emitting regions.

The Ca~II lines in the spectrum are interesting. The IR triplet lines at
$\wll$8498, 8542, 8662 are clearly present, while for the H\&K lines and
the forbidden $\wll$7291.5, 7323.9 lines we can only set upper limits. Fransson
\& Chevalier (1989) and Ferland \& Persson (1989) have discussed the relative
strengths of these lines as a function of density and temperature. Using their
results we find that the fact that $I(\wll$7291.5, 7323.9)/$I(\wll$8498,
8542, 8662) $< 0.25$ and $I$(H\&K)/$I(\wll$8498, 8542, 8662) $< 0.12$
means that the density must be $\gsim 10^9 \cmc$. This result is only weakly
sensitive to the temperature.

The parts of the spectrum which are not well fitted by the model can in most
cases be identified with regions of strong Fe~II emission. This includes the UV
between $2300 - 3000$ \AA, some regions in the range 4200~\AA\ to 5300~\AA, the
$\wl$7155 line, and the fluorescence lines above $\sim 8000$~\AA\ (see below).

In Figure \ref{fig6} we show the line profiles and fits of the H$\alpha$,
[O~I], [O~II], and [O~III] lines to illustrate the fits for our assumed radial
emissivity model. The figure also demonstrates the complex spectrum which
results from the blending of the individually fairly simple line profiles.

Turning now to the other metal lines coming from the intermediate-velocity
component, the UV line-intensity ratios require N/C $= 2.5 -5$ and O/N $= 3.5
-4.7$, by number.  The high N/C ratio, in combination with the relatively large
O/N ratio, strongly indicates incomplete CNO processing in this component.

While the profiles of the high-ionization C, N, O, and Si lines in the far-UV
are consistent with the same shell-like emissivity as the oxygen lines, this
emissivity model shows a bad fit of the Mg~II $\wll$2796, 2803 line. The peaked
profile shows that this line is instead likely to originate in the H and
He-rich broad-line component, rather than in the metal-rich component.

\subsection{Fe~II -- Ly$\alpha$ fluorescence}
\label{sec_fluor}

Based on the Fe~II -- Ly$\alpha$ fluorescence calculation by Sigut \& Pradhan
(1998a,b), we have modeled the near-IR spectrum of SN 1995N. We have used the
same model as in Figure \ref{fig5b}, but now added the Fe~II lines, with
wavelengths and fluxes from the list of Sigut \& Pradhan. The line profiles
were assumed to be the same as that of H$\alpha$.  In the upper panel of Figure
\ref{fig4b} we show the resulting spectrum, while in the lower panel we show
for comparison the spectrum without Fe~II lines. The observations are from the
day 1769 VLT spectrum, but the day 716 and 1037 Lick spectra show the same
features. The day 1007 Keck spectrum only extends to $\sim 9000$~\AA.

As can be seen, there is a striking improvement with the observed features
between 8000~\AA\ and 9500~\AA\ when the Fe~II lines are included. In
particular, the feature at $\sim 9150$~\AA\ can be identified with the Fe~II
$\wll$9071, 9128, 9177 lines, and provides a good fit both in wavelength and
profile of the blend.

One of the strongest expected fluorescence-pumped Fe~II lines is at
$\lambda$8451, and there is indeed a strong feature at this
wavelength.  Another possible identification for this feature is the
recombination line of O~I $\lambda$8446.  This line is normally
accompanied by O~I $\lambda$7774, which is not present in the broad
component (although it is visible in the intermediate component; see
\S~\ref{sec_result}).  The recombination contribution to the $\wl$8446
line is expected to be close to the $\lambda$7774 line, and is
therefore insufficient to explain this strong feature.  The
$\lambda$8446 line can be produced by fluorescence through the
Ly$\beta$ line (Grandi 1980).  Leibundgut et al. (1991) identified the
O~I $\lambda$8446 line in the spectrum of SN 1986J. The Ly$\beta$ --
O~I $\lambda$8446 fluorescence also results in a strengthened O~I
$\wl$1300 multiplet. Due to its narrowness, however, this is likely to
be absorbed by the O~I in the host galaxy.  In our Keck day 1007
spectrum of SN 1995N, the center of the emission feature gives a
better match to Fe~II $\lambda$8451 than to O~I $\lambda$8446. An
additional argument in favor of this mechanism is that Fe~II
$\wll$2506, 2508, which is indicative of Ly$\alpha$ pumping, is the
strongest Fe~II feature in the UV. This has previously been seen in
$\eta$ Carina, RR Tel, and KQ Pup (Baratta, Cassatella, \& Viotti
1995; Johansson \& Jordan 1984; Redfors \& Johansson 2000).  Other
lines in the UV expected to be enhanced by fluorescence are the Fe~II
multiplets 399, 391, and 380 at 2850~\AA, multiplet 373 at 2770~\AA,
and multiplet 363 at 2530~\AA. These most likely contribute to the
broad wings of the Mg~II lines, but are too blended to be uniquely
identified. IR lines at $\sim 1.7~\mu$m are also expected to be strong
(Johansson \& Hamann 1993), but are outside our observed wavelength
range.

Given that the model was intended for conditions in an AGN, the fit is
surprisingly good. Only at $\gsim 9300$~\AA\ are there a few features (such as
at $\sim 9573$~\AA) which are not well reproduced, but the noise level at these
wavelengths is considerable. One should also note that the relative strengths
depend on the uncertain atomic data, as well as on the exact Ly$\alpha$
profile. In particular, the larger line width in SN 1995N compared with those
assumed in the AGN calculation by Sigut \& Pradhan (1998a,b) model can change
the results appreciably. Sigut \& Pradhan assumed a static Ly$\alpha$ profile
with only damping included. An expanding medium means that the Ly$\alpha$ line
can feed many more fluorescence channels, which is likely to enhance the
fluorescence effects.

Finally, we note the strong He~I $\wl$10830 line in the day 1769
spectrum. Although the region beyond $\sim 9300$~\AA\ becomes increasingly
noisy, this feature clearly stands out above the noise level. Compared to the
strength expected by pure recombination, it is enhanced by collisional
excitation by a factor of $\sim 8$, which is common in many objects (e.g.,
Osterbrock 1989, p. 111).

\section{TIME EVOLUTION}
\label{sec_time}

In Figure \ref{fig6a} we show the evolution of the reddening-corrected
H$\alpha$ luminosity with time on a linear-log scale. We see that the
evolution can be well approximated with an exponential decay,
$L({\rm H}\alpha) \propto \exp(-t/\tau)$, where $\tau = 460$ days. This
is much longer than the ${}^{56}$Co decay time scale, and we conclude
that the circumstellar interaction is responsible for the longevity, although
there is no obvious explanation for this particular time dependence.

An efficient way of displaying the change of the spectrum with time is
to plot the ratio of two spectra as a function of wavelength. In
Figure \ref{fig7a} we show a running average taken over 2~\AA\ of the
day 1799 VLT and day 1007 Keck spectra.  The presence of two separate
components in the spectrum is apparent. While most of the spectrum has
decreased by a factor of $\sim 6$, the intermediate-component [O~I],
[O~II], and [O~III] lines only decreased by a factor of 2--3.  The O~I
$\wl$7774 line shows a similar decrease, while the $\wl$8450 feature
follows the H~I lines, consistent with it being mainly due to Fe~II
fluorescence lines. In addition, Mg~I] $\wl$4571 belongs to the class 
of slowly declining lines.

Relative to H$\beta$, the peak fluxes of the [O~I] and [O~II] lines increased
by a factor of $\sim 3$ between 1997 and 1999, while the [O~III] line increased by a
more modest factor of $\sim 1.6$ between these observations. The [O~III]
$\wll$4959, 5007/$\wl$4363 ratio remained nearly constant at $\sim 7.9$, while
the O~I $\wl$7774/[O~II] $\wll$7320, 7330 ratio decreased from $\sim 0.43$ to
$\sim 0.20$. Repeating the analysis in \S~\ref{sec_interm} (Fig. \ref{fig4a}),
we find that the temperature in the oxygen-rich gas had not changed much, while
the electron density decreased from $\sim 3 \times 10^8 \cmc$ to $\sim 1 \times
10^8 \cmc$. This is less than expected from a simple $n_e \propto t^{-3}$ law,
but our analysis assumes that the relative sizes of the O~II and O~III zones
remain constant, and that the temperatures and densities in these two zones are
similar. Neither of these conditions is obviously satisfied.  In addition, the
supernova may be older than our estimate, and the uncertainties in the relative
fluxes are considerable. Nevertheless, the behavior is qualitatively that
expected for expanding ejecta.

In Figure \ref {fig7} we show the relative change of H$\alpha$, [O~I]
$\wll$6300, 6364, [O~II] $\wll$7320, 7330, and [O~III] $\wll$4959, 5007 on a
velocity scale. While the [O~I] and [O~II] line profiles changed by roughly the
same factor over the whole velocity range, $-5000$ to $+5000 \kms$, the red
side of the [O~III] line decreased by a large factor relative to the blue.  The
most conceivable reason for the rapid [O~III] change is that blending with
another line, decreasing similarly to the H~I and He~I component, causes the
apparent decrease in the red wing. The most likely candidates for this are the He~I
$\wll$5015, 5049 and Fe~II lines.  The blue side behaved in a manner similar to
the [O~I] and [O~II] lines.  We therefore conclude that there is little change
in the overall profiles of the oxygen lines. When comparing the width of the
best-isolated [O~II] line, we find a nearly constant blue extension, while the
width on the red side decreased by $\sim 450 \kms$ during this period.

The evolution of the H$\alpha$ line profile differs dramatically from that of
the oxygen lines.  As shown in the upper-left panel of Figure \ref {fig7}, the
red wing up to $\sim 3000 \kms$ decreased by almost a factor of 4 relative to
the blue, with a maximum decrease at $\sim 1100 \kms$.  To further illustrate
the different evolution of the wings of H$\alpha$, we have reflected the blue
component in the spectra from 1996 -- 1999 (Fig. \ref{fighaevol}), so that the
blue and red wings can be compared directly. The fluxes have been normalized to
the same peak value. We see that the day 716 Lick spectrum (dashed lines) showed
nearly identical red and blue wings. The day 1007 Keck spectrum (solid lines) shows
a clear evolution in that for a given velocity the red wing decreased faster
than the blue wing. This is even more pronounced in the 1999 VLT spectrum,
where at $\sim 1500 \kms$ the flux of the red wing is only $\sim 20$\% that of
the blue wing at the same velocity. Equivalently, for a given intensity the red
wing has a much smaller extension.  For example, at $\sim 10$\% of the peak
flux the blue wing extends to $\sim 1900 \kms$, while the red wing only extends
to $\sim 1080 \kms$. Relative to the peak flux, the blue wing shows a steady
increase in flux with time, while the red wing displays the opposite
trend. Similar evolution is seen for both H$\beta$ and He~I $\wl$5876, although
the continuum contribution and line blending make it difficult to compare
directly to H$\alpha$.  Baird et al. (1998) have presented evidence that the
line slowly narrows with time.  We can therefore confirm this trend to the
latest epochs.  The H$\alpha$/H$\beta$ ratio in the 1999 spectra decreased
somewhat relative to the 1997 and 1996 spectra. This is consistent with the
decreasing density during this time interval.

The intensity ratio of the He~I lines to the Balmer lines does not change
much. The two He~I lines least affected by blending, He~I $\wll$5876, 7065,
have almost the same ratio relative to H$\beta$ at all epochs, and are all
reproduced with a relative He~II/H~II abundance of $\sim 0.35$ by
number. Together with the similar line profiles, this argues for the fact that
the He~I and Balmer lines arise in the same region. Their relative flux should
therefore reflect their abundance ratio.

In most cases the fluxes of the narrow lines are uncertain, especially in the
day 1799 spectrum. For the strongest lines, like [O~III] $\wl$5007 and [Fe
VII] $\wl$6087, the decrease is by a factor of 5--10, with no systematic
distinction between lines from different elements.  Although the flux of the 
[O~III] $\wl 4363$ line is especially uncertain due to the underlying broad
component, there is an indication for an increase in the [O~III]
$\wl$5007/$\wl$4363 ratio from $\sim 4.1$ to $\sim 6.6$. However, because this
ratio depends on both density and temperature, one cannot directly translate
this into a unique change in either of these parameters.

\section{DISCUSSION}
\label{sec_discuss}

\subsection{Nature of the Emission Regions}
\label{sec_nature}

The general scenario for the emission from SN 1995N and related supernovae
(generally Type IIn) is
interaction of the supernova ejecta with circumstellar gas.  Our spectra of SN
1995N show that the interaction is complex, but that there is sufficient
diagnostic information on the emission components to deduce some of the
interaction details.  All the emission components are likely to be heated and
ionized by the energetic radiation from the interaction region.

\subsubsection{Narrow Lines}
\label{sec_narrow_orig}

Based on the unresolved line widths, the density of $\sim 2 \times 10^6 \cmc$,
and the low filling factor, the narrow-line emission can be identified with
clumps in the preshock circumstellar gas. The velocity is likely to be $< 50
\kms$, typical of the circumstellar media of red supergiants.

\subsubsection{Intermediate Component}
\label{sec_interm_orig}

The intermediate-velocity component is identified with unshocked supernova
ejecta.  The maximum velocity in this component suggests either that the
reverse shock front has a minimum velocity of $\sim 5,000 \kms$, or
alternatively that the emission is coming from a component which is
photoionized by the shock but only extends in abundance to this velocity.  The
fact that this component is only seen as dense, oxygen-rich gas makes it
natural to identify it as processed gas from the core. Assuming that the oxygen
core expands at $5,000 \kms$, the typical oxygen density will be
\begin{equation}
n_O \approx 4.1\times 10^5 f^{-1} {\left( M_{\rm core} \over 2 \Msun \right)}
{\left( V_{\rm core} \over 5000 \kms \right)}^{-3}
{\left( t \over 1000 \rm ~days \right)}^{-3} ~\cmc,
\label{eq_b}
\end{equation}
where $f$ is the filling factor of this component in the core, and we have
assumed an oxygen-dominated composition. With $f \approx 0.1$, as was the case
for SN 1987A (e.g., Spyromilio \& Pinto 1991), the density is of the same order
as found from the [O~III] and similar lines. We note that the density deduced
from the models is based on the assumption that the temperatures and densities
are the same in the O~I, O~II, and O~III zones.

A velocity of $\sim 5,000 \kms$ is high for the oxygen core of a Type IIP or
Type IIL supernova, but in the range found in Type Ib/Ic supernovae, as well as in
the Type IIb SN 1993J (e.g., Houck \& Fransson 1996; Matheson \etal 2000). This
is an important result because it is consistent with a picture where the
progenitor has lost most of its hydrogen-rich envelope before the explosion.

\subsubsection{Broad Component}
\label{sec_broad_orig}

While most of the hydrogen-rich gas comes from gas moving at velocities of
$\lsim 2000 \kms$, there is a smooth extension to $\sim 10^4 \kms$. Any model
must explain the wide range of velocities for this component. In particular,
the fact that there appears to be hydrogen-rich, cool gas moving at up to
$10,000 \kms$ implies that there are some regions where the shock region
extends considerably beyond the reverse shock front.



To explain this, one may identify three main alternatives for the broad,
hydrogen-rich component. The first possibility is that it comes primarily from
unshocked, freely expanding ejecta, extending from low-velocity hydrogen in the
core up to the maximum ejecta velocity. A second possibility is that it comes
from shocked ejecta gas, behind the reverse shock. The final possibility
is that it comes from shocked circumstellar gas, either in the form of clumps
or a more uniform medium.  Our synthetic spectral models in \S~4 are
appropriate for the first possibility.

Before discussing these possibilities, we note that the Ca~II lines, the
absence of forbidden lines, and the H$\alpha$/H$\beta$ ratio indicate a density
$\gsim 10^9 \cmc$ for this component (\S~\ref{sec_result}).  A further
independent lower limit to the density can be obtained by considering the state
of ionization in this gas. The luminosity in H$\alpha$ was $\sim 2.3 \times
10^{40} \ergs$ on day 1007 (\S~\ref{sec_broad}). Typically, the conversion
efficiency from X-rays to H$\alpha$ is $\sim 1$\% (CF94), indicating a total
X-ray luminosity of $2 \times1 0^{42} \ergs$. This is much more than was
observed by {\it ROSAT}
(Fox et al. 2000), but as calculations for SN 1993J showed (Fransson,
Lundqvist, \& Chevalier 1996), almost all the X-ray emission can be absorbed in the
ejecta and shocked gas, and only the hard X-rays escape the supernova. With
this luminosity the ionization parameter on day 1007 is $\xi \approx 1 \times
10^2 (n/10^6 \cmc)^{-1}$.  For a uniform density, hydrogen-dominated envelope
expanding at $\sim 10,000 \kms$, the density is
\begin{equation}
n \approx 4.1\times 10^5 f^{-1} {\left( M_{\rm env} \over 1 \Msun \right)}
{\left( V_{\rm env} \over 10,000 \kms \right)}^{-3}
{\left( t \over 1000 \rm ~days \right)}^{-3} ~\cmc.
\label{eq_c}
\end{equation}
Therefore, if the emission came from unshocked ejecta and $f~ \gsim ~0.1$, the
ionization parameter would be very high, $\xi ~\gsim ~10$, and one would expect
strong emission from ions like [O~III] from this component, contrary to the
observations. Instead, it is dominated by Mg~II and Fe~II emission, in addition
to the H~I and He~I lines, characteristic of a partially ionized, high-density
region (e.g., CF94). It is therefore unlikely to come from a uniform-ejecta
component, and only if clumping of this component is very high ($f~ \lsim ~0.01$)
is an ejecta origin likely. We therefore think the first possibility 
given above is
unlikely.

The density in the broad component is much higher than that in the narrow-line
component, but the temperature is expected to be only slightly smaller.  The
conclusion is that the pressure in the gas is relatively high and one
possibility is shocked circumstellar or ejecta gas.  The gas pressure is $p/k =
2 \times 10^{13} n_9 T_4\cmc$ K, where $n_9$ is the H density in units of $10^9
\cmc$ and $T_4$ is the temperature in units of $10^4$ K.  The pressure
generated by a shock front moving into gas with a density of $10^6\cmc$ is $p/k
= \rho_o v_s^2/k = 1.5 \times 10^{14} v_{s3}^2\cmc$ K, where $v_{s3}$ is the
shock velocity in units of $10^3 \kms$.  If the emission is from shocks moving
into a medium with a density of this order, the required high density can
easily be produced.

We can identify two possibilities for such a shocked gas: it could be either
clumps caused by instabilities in the shocked ejecta behind the reverse shock,
or clumpiness and instabilities of the shocked circumstellar medium. This is
similar to the two scenarios discussed by Chugai \& Danziger (1994) for SN
1988Z.

\textit{Shocked ejecta:} This is a version of the scenario discussed for
``normal'' interacting supernovae like SN 1979C or SN 1993J (CF94; Fransson
et al. 1996). The fact that the line profiles in SN 1995N are
peaked as opposed to the boxy profiles seen in SN 1993J, however, makes it
necessary to modify this scenario. One possibility is that the circumstellar
medium is highly asymmetric, as in the picture discussed by Blondin, Lundqvist,
\& Chevalier (1996). Indications for a highly asymmetric circumstellar medium
are seen in some red supergiants in a superwind phase (see
\S~\ref{sec_progenitor}).  Most of the hydrogen emission could in this case
come from the cool, dense shell expected behind the reverse shock (CF94). The
shape of the H$\alpha$ line profile then reflects the decreasing emission
measure as one moves away from the equator. The maximum ejecta velocity is
likely to occur at the poles, while the strongest interaction, with most of the
energy release, occurs in the high-density circumstellar gas at the
equator. The deceleration of the ejecta will therefore be most rapid here. If
the hydrogen envelope mass is low, the reverse shock at the equator may rapidly
recede to the massive oxygen core.  This could qualitatively explain the
presence of the high-ionization, oxygen-rich shell observed in the
intermediate-component line profiles. The hydrogen-rich gas in the cool shell
behind the reverse shock could then be excited by the X-rays from the shock,
giving rise to an extended partially ionized zone. A prediction is that the
maximum velocity of the shocks, and therefore the line widths, should decrease
with time.

This general scenario for SN 1995N receives support from observations of the
related supernova SN 1986J in NGC 891, which had a strong H$\alpha$ line with
FWHM $\approx 700 \kms$ and oxygen lines with widths of several $10^3 \kms$
(Leibundgut et al. 1991).  It was a strong radio source, and VLBI images showed
a shell with protrusions extending at least a factor of two beyond the shell
radius (Bartel et al. 1989, 1991).  Images on different dates revealed
expansion; the outer protrusions are moving at $\sim 15,000 \kms$, while the
shell is expanding at $\sim 5,000 \kms$.  These images suggest a picture that
is consistent with that deduced from the spectra of SN 1995N.

The origin of the protrusions is uncertain, but there are computational
simulations that suggest that they could arise from asymmetries or
inhomogeneities in the circumstellar medium.  Blondin et al.
(1996) found that supernova expansion into a circumstellar medium with a
density deficit in the polar direction could produce polar protrusions with a
radius 2--4 times that of the main shell.  The problem in comparison to the
images of SN 1986J is that there are more than two protrusions.  Jun, Jones, \&
Norman (1996) carried out a simulation of the interaction of a supernova with a
clumpy medium and found that the Rayleigh-Taylor instability that develops
where the ejecta are decelerated by the surrounding medium is accentuated by
the presence of clumps.  The Rayleigh-Taylor fingers reach the outer shock
front.  The protrusions in the simulation did not reach the extent required for
SN 1986J, but the simulation was run for a limited amount of time and the
density of the surrounding medium was uniform.  The circumstellar wind case
remains to be investigated.

\textit{Shocked circumstellar clouds:} In this scenario, the emission comes from
the cool gas behind the slow radiative shocks propagating into the same clumpy
gas clouds as are responsible for the narrow-line emission.  The blast wave may
then propagate in the intercloud medium, while the high-density circumstellar
clumps are left behind. The maximum ejecta velocity will decrease only slowly,
while the shocked clouds will be accelerated on a time scale which depends on
the column density of the clouds. As simulations show (e.g., Klein, McKee, \&
Colella 1994; Jun et al. 1996), however, the cloud may gradually be
evaporated, and finally be dissolved and merge with the ejecta gas. In this
scenario one expects a smooth distribution of gas, with velocities ranging from
newly shocked clouds with low velocity, or clouds with very high column
density, to gas which has been accelerated to the ejecta velocity. This agrees
qualitatively with the observed H$\alpha$ line profile. The observed gradual
increase of the flux in the blue wing with time is also consistent with this
scenario. The decrease in the red wing, however, is not naturally explained,
unless it is caused by dust absorption, either from pre-existing dust in the
circumstellar gas, or from newly formed dust in the ejecta, as discussed for SN
1998S by Gerardy \etal (2000).

\subsubsection{Location of the Energy Source}
\label{sec_energy_orig}

An important question is the location of the energy source for the excitation
of the gas, and there are several possibilities. First, the source could be the
shock separating the high-velocity hydrogen ejecta from the circumstellar
gas. In the uniform-medium case this will be at the highest velocity and also
maximum ejecta radius. This is the model discussed by CF94, and applied to SN
1993J by Fransson et al. (1996).  However, it is difficult to
understand the presence of the high-ionization, oxygen-rich shell at only $\sim
5000 \kms$, since most of the X-rays will be absorbed by the unprocessed ejecta
between the shock and the oxygen core.

In the clumped circumstellar medium case the bow shocks may be distributed over
a large radius, and immersed in the expanding ejecta, depending on the time of
shocking and their column densities.  The oxygen-shell emission is again
difficult to explain. It may, however, explain the distributed nature of the
hydrogen emission.

\subsection{Type IIn Progenitors}
\label{sec_progenitor}

Some clues to realistic scenarios for the structure of the circumstellar medium
can be obtained from observations of Galactic supergiants in very advanced
evolutionary stages. While most red supergiants, like $\alpha$ Ori, have fairly
modest winds with mass-loss rates $\lsim 10^{-5} \mll$, there is a handful of
massive stars having more extreme mass-loss rates, $10^{-4} -10^{-3} \mll$,
often referred to as ``superwinds.'' Van Loon \etal (1999) find that two out of
eight red supergiants in an LMC sample have mass-loss rates $\gsim 10^{-3}
\mll$, while the others have moderate mass-loss rates $\lsim 10^{-5} \mll$.

The best-studied superwind cases are VY CMa and IRC +10420. The nature of the
superwind phase is not yet clear, but can be related either to a specific
evolutionary phase, occurring for most red supergiants, or to a phase only
occurring for a specific mass range. A hint that the latter is the case comes
from the fact that both VY CMa and IRC +10420 are very luminous stars, with $L
\approx (2-5)\EE5 {~\rm L}_\sun$.  From evolutionary models Wittkowski, Langer,
\& Weigelt (1998) estimate a zero-age main sequence (ZAMS) mass of 30--40
$\Msun$ for VY CMa.  The mass-loss rate is determined to be $\sim (2-3) \times
10^{-4} \mll$ for a wind velocity of $39 \kms$ (Danchi \etal 1994).

Several authors have found an asymmetric distribution of the circumstellar gas
and dust around VY CMa.  From {\it HST} imaging Kastner \& Weintraub (1998)
find an asymmetric core region with wavelength dependent FWHM $\approx 10^{15}$
cm in the optical. The star itself has a radius of $\sim 1.2 \times 10^{14}$
cm. Outside of this is an asymmetric reflection nebula with an extent of $\sim
6 \times 10^{16}$ cm, with the core close to its edge. Based on symmetry
arguments, Kastner \& Weintraub speculate that this is only one half of a
bipolar nebula, with the other half obscured by an opaque equatorial dust
disk. The same conclusion is reached by Monnier \etal (1999) from
high-resolution observations of the core and extended structure.

The supergiant IRC +10420 is similar to VY CMa in terms of mass-loss
rate and mass, but probably in an even later evolutionary state,
intermediate between a red supergiant and a Wolf-Rayet star. 
There is also evidence  for a disk-like geometry in this case, based on line
profiles (Jones \etal 1993) and on {\it HST} and IR imaging (Humphreys \etal
1997). 

Wittkowski et al. (1998) propose that the asymmetric mass loss
arises as a result of rapid rotation in the phase preceding the evolution off
the Hayashi line when the major part of the envelope is lost. VY CMa is now
estimated to have a mass of $\sim 15 \Msun$, meaning that it has lost $\sim 25
\Msun$ to the circumstellar medium. The helium fraction at this phase is $\sim
0.4$ by mass. This is lower than, although qualitatively consistent with, our
results for SN 1995N, but is obviously sensitive to mixing prescriptions, the
exact evolutionary status, and other factors.  As an alternative, Soker (2000)
proposes that the asymmetric distribution is caused by the concentration of
magnetic spots in the equatorial plane, which through the lower photospheric
temperature causes dust formation preferentially at the equator.

Based on these cases, it seems probable that SN 1995N, and other Type
IIn supernovae, arise from progenitors similar to VY CMa or IRC
+10420. This is consistent with the high mass-loss rates needed to
explain the strong interaction, as reflected in the total
luminosity. A low hydrogen envelope mass is needed to explain the
large oxygen core velocity. The observed high He/H ratio is also
consistent with high mass loss.  An asymmetric circumstellar
distribution will result in faster expansion of the supernova ejecta
in the polar directions, while it will be slowed down by the
equatorial disk or torus as modeled by Blondin et al. (1996). The
radius of the inner boundary of the dusty disk in VY CMa is estimated
to be $\sim 10^{15}$ cm (Monnier \etal 1999). If the density in the
disk is large enough, the radius of the inner edge would not be much
affected by the supernova ejecta. The relative size of the ejecta and
obscuring disk would therefore increase with time. If the disk is
opaque due to dust, an increasingly large fraction of the ejecta on
the far side of the supernova would be obscured. This could explain
the fact that the red side of the H$\alpha$ line decreased
considerably faster than the blue side. The exact temporal evolution
depends on the degree of asymmetry of the ejecta.

Evidence for dusty environments comes from several Type II
supernovae. Especially interesting are the observations of SN 1998S, where
Gerardy \etal (2000) find strong evidence for dust emission. As Gerardy \etal
discuss, although this dust could in principle be formed in the ejecta, perhaps
a more likely interpretation is pre-existing circumstellar dust, heated by the
supernova (see also Fassia et al. 2001).

   From observations of H$_2$O masers in VY CMa, Richards, Yates, \& Cohen (1998)
estimate a density of $\sim 5 \times 10^9 \cmc$ in the masing regions, while
other species give a density of $\sim 10^8 \cmc$, indicating a considerable
clumpiness of the circumstellar medium. One then speculates that these
masing regions could be related to the high-density clumps needed to decelerate
the supernova blast wave. 

Finally, the small fraction of supernovae in this class, $\lsim 5$\%
(Cappellaro et al. 1993), is at least qualitatively consistent with them coming
from a fairly small fraction of the high-mass stars.

\section{CONCLUSIONS}
\label{sec_concl}

Late-time spectral observations of SN 1995N are presented, covering the UV to
near-IR regions. Three kinematic components are found, which most likely are
powered by X-rays from the interaction of the ejecta and the circumstellar
medium of the progenitor. Spectral modeling shows evidence of an increased
helium abundance, as well as CNO burning. The first evidence for Ly$\alpha$
pumped fluorescence of Fe~II lines in supernovae is found. It is likely that
other supernovae interacting with their circumstellar media show similar
signatures, especially in the near-IR. While the narrow lines come from
unshocked circumstellar gas and the intermediate-velocity component from
processed ejecta, the high-velocity unprocessed gas has a more uncertain
origin. Different possibilities based on either clumpy or asymmetric
circumstellar media are discussed, without strongly favoring either of these
possibilities. The temporal evolution of the processed and unprocessed gas is
distinctly different, again emphasizing the different origins of these
components. We propose that the progenitors of the narrow-line supernovae (Type
IIn) are similar to red supergiants in their superwind phase, as most of the
hydrogen-rich gas is expelled in the last $\sim 10^4$ years before
explosion. This explains the high oxygen core velocity. We also present
evidence that the circumstellar medium is either asymmetric or clumpy.

\acknowledgments

We are grateful to Anil Pradhan and Aron Sigut for sending us their
fluorescence results in detailed form.  Financial support for this work was
provided to A.V.F.'s group by NSF grants AST-9417213 and AST-9987438, by the
Guggenheim Foundation, and by NASA through grants GO-6043, GO-6584, and GO-8602
from the Space Telescope Science Institute, which is operated by AURA, Inc.,
under NASA contract NAS 5-26555. Support was also provided to R.A.C. by NASA
grant NAG5-8232, and to C.F.'s group by the Swedish Space Board and Swedish
Research Council.  Some of the data presented herein were obtained at the
W. M. Keck Observatory, which is operated as a scientific partnership among the
California Institute of Technology, the University of California, and NASA; the
Observatory was made possible by the generous financial support of the
W. M. Keck Foundation. We thank the staffs at the {\it HST}, ESO, Lick, and
Keck Observatories for their overall assistance, as well as Andrea Gilbert,
Jeff Newman, and Adam Riess for their help with some of the observations.

\clearpage

\begin{figure}  
\plotone{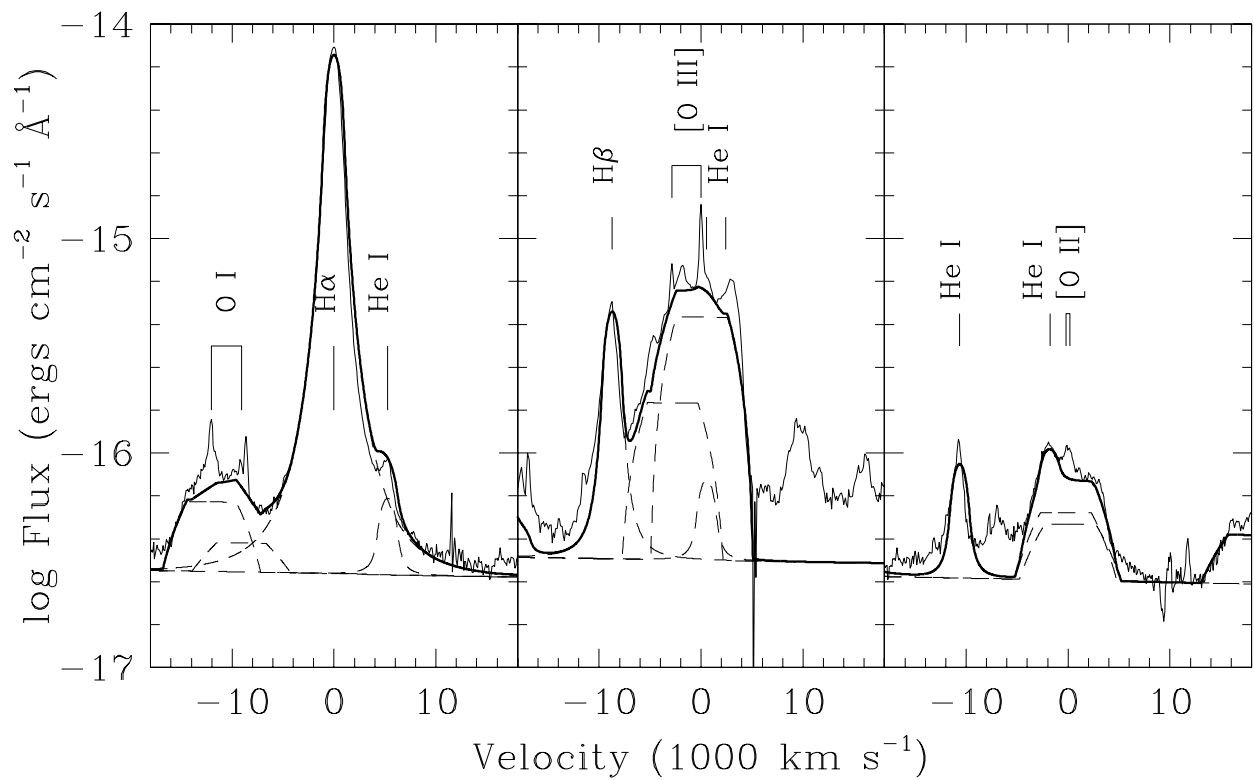}
\caption{Spectral fit and decomposition of the H$\alpha$, [O~I], [O~II], and
[O~III] line profiles from the day 1007 spectrum. The velocity
scale is counted from H$\alpha$, [O~III] $\wl$5007, and [O~II] $\wl$7325, 
respectively. The dashed lines give the most important separate
contributions to the total line profile. The narrow components of
the [O~I], [O~II], and [O~III] lines are apparent as the excess
emission.}
\label{fig6}
\end{figure}

\begin{figure}  
\plotone{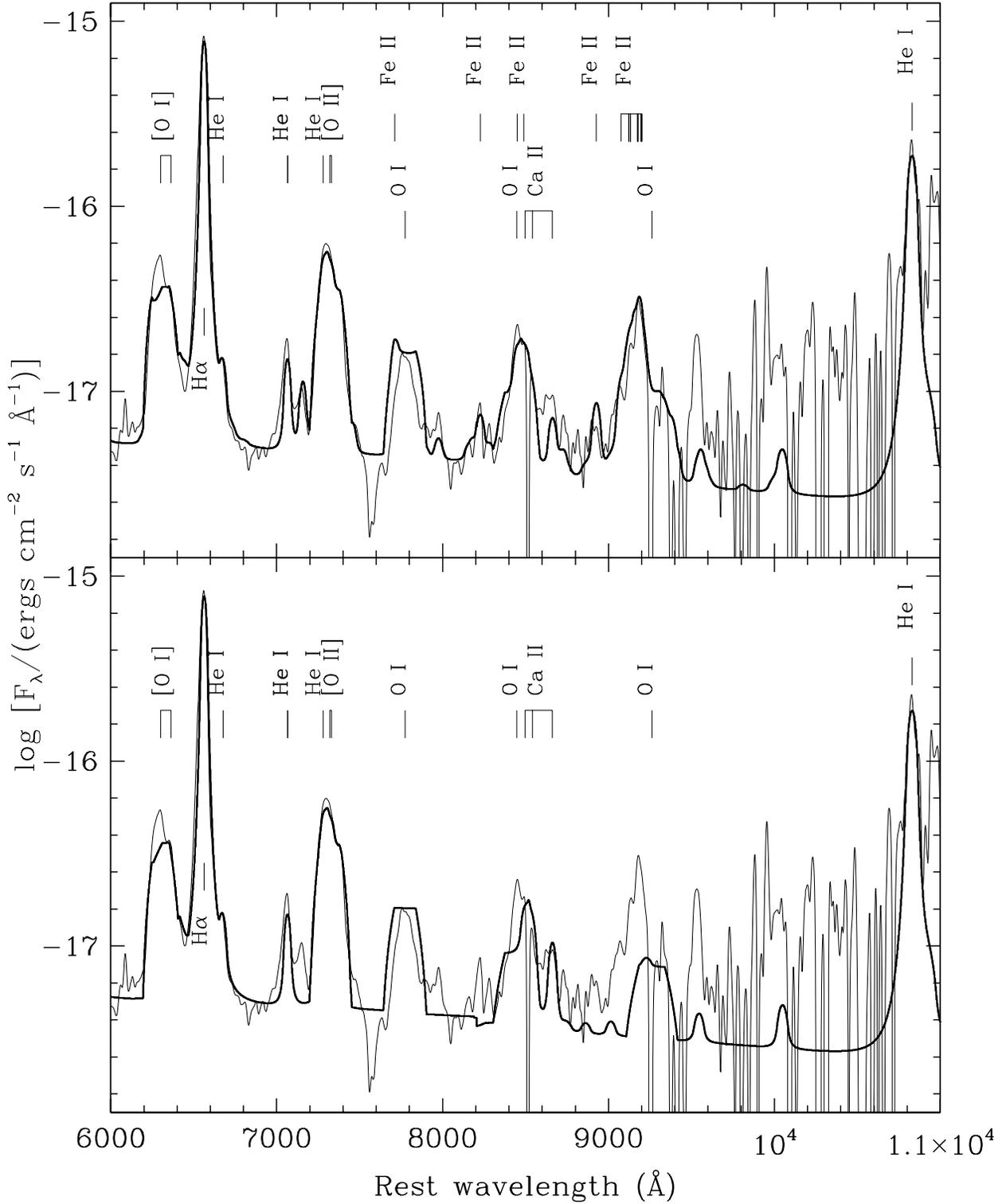}
\caption{The observed near-IR region of the spectrum of SN 1995N on day 1769,
together with the synthetic spectrum expected based on the Ly$\alpha$
fluorescence mechanism (thick line). The upper panel includes the
Ly$\alpha$ fluorescence, while the lower does not. Note especially the
greatly improved fits at $\sim 8500$~\AA\ and $\sim 9150$~\AA. Also
note the strong He~I $\wl$10830 line.}
\label{fig4b}
\end{figure}

\begin{figure}  
\plotone{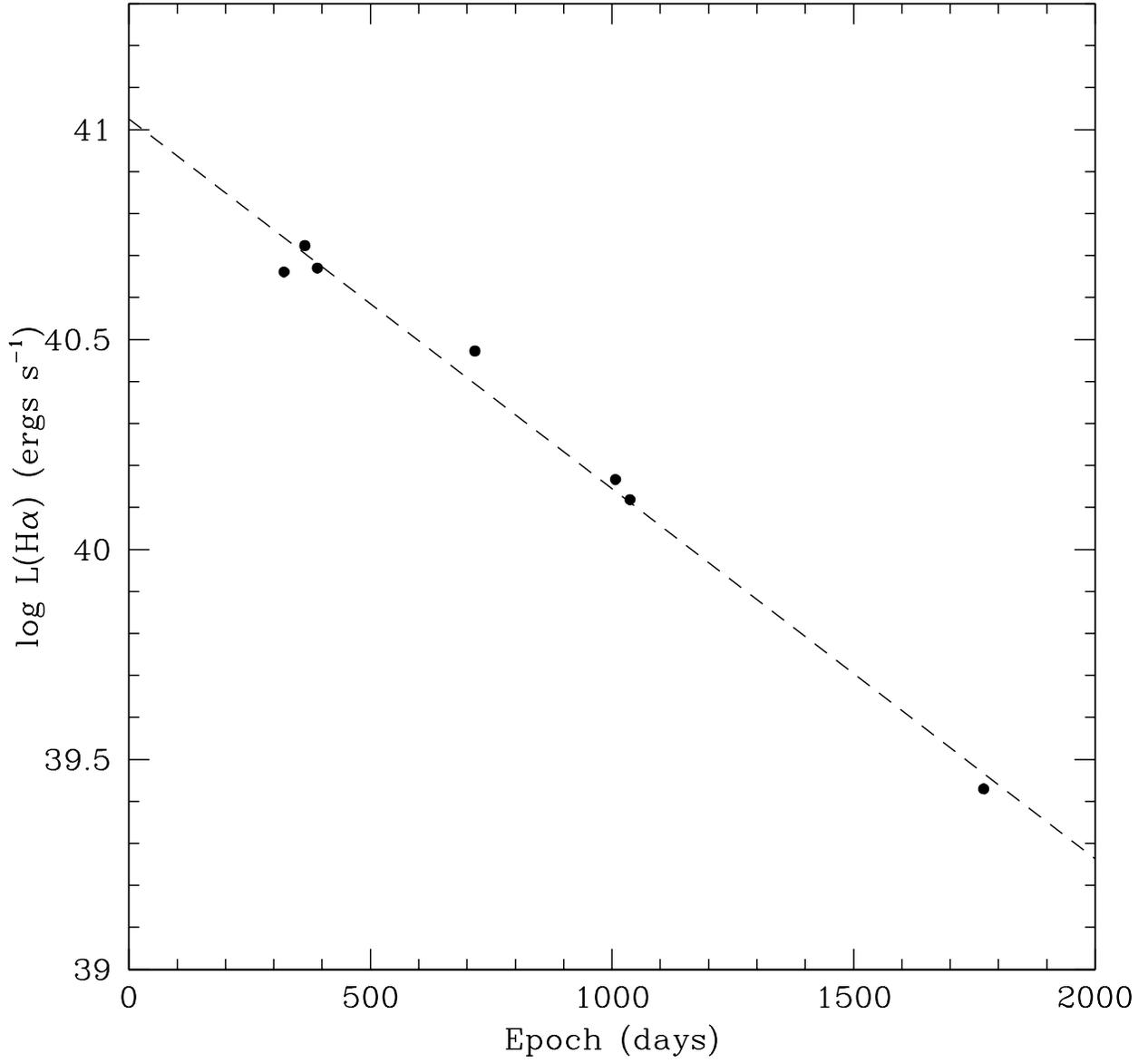}
\caption{Evolution of the H$\alpha$ luminosity with time. The dashed line 
gives a least-squares fit to the data.}
\label{fig6a}
\end{figure}

\begin{figure}  
\plotone{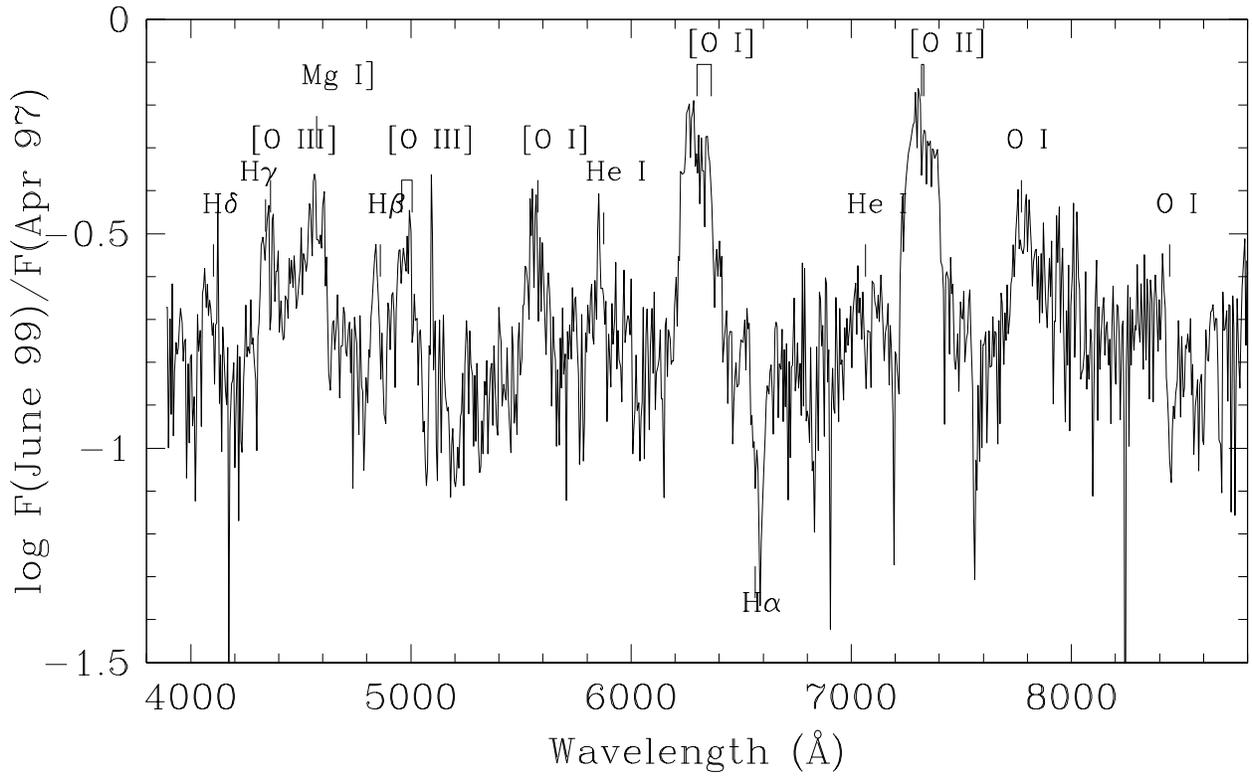}
\caption{Ratio of the day 1799 and day 1007 spectra with
identifications of the most important lines. Note the peaks of the [O~I], [O~II],
[O~III], and Mg~I lines, showing that these belong to a component
separate from the H~I, He~I, and Fe~II lines.}
\label{fig7a}
\end{figure}

\begin{figure}  
\plotone{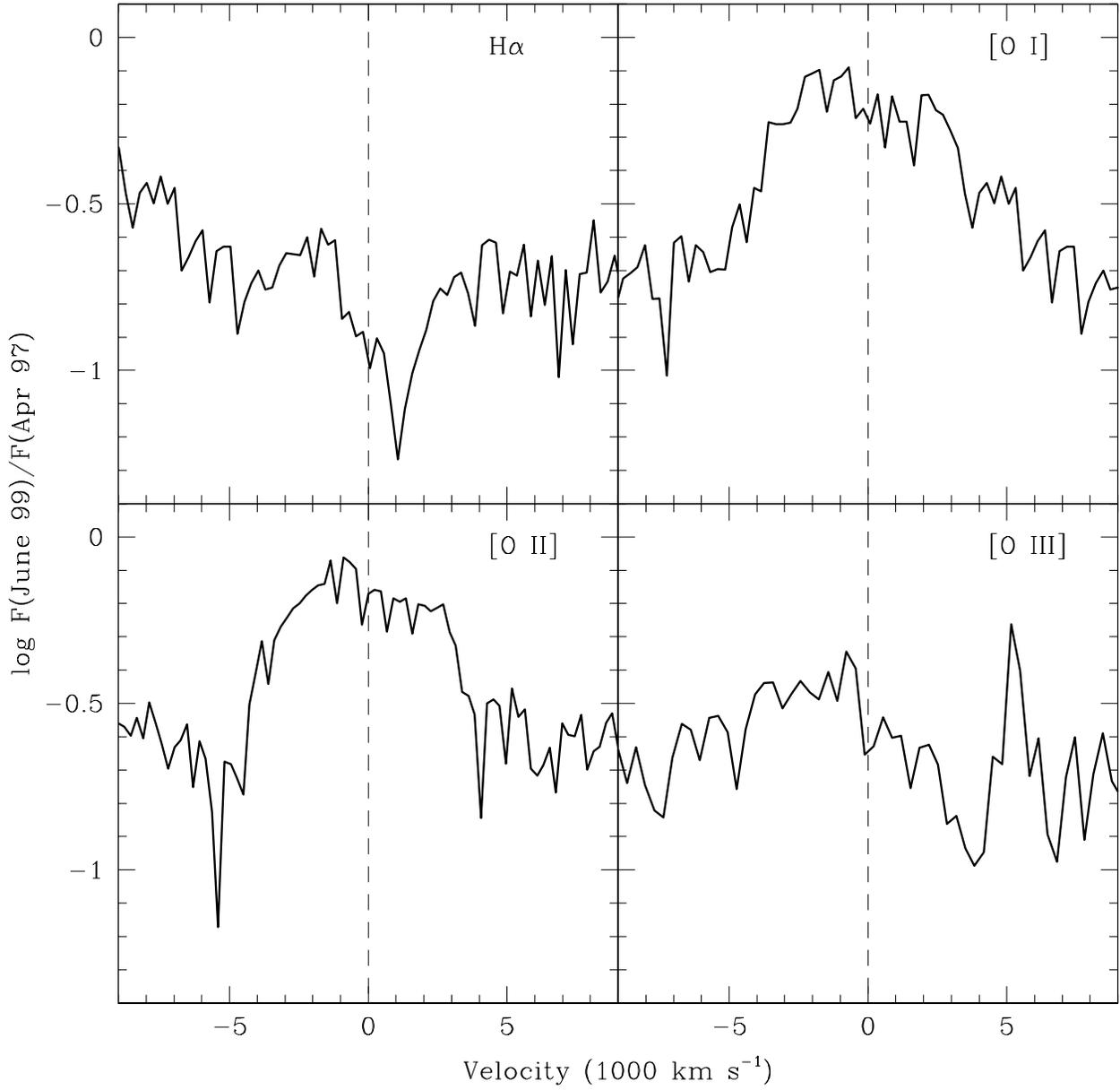}
\caption{Ratio of the day 1799 and day 1007 spectra for
H$\alpha$, [O~I], [O~II], and [O~III].}
\label{fig7}
\end{figure}

\begin{figure}  
\plotone{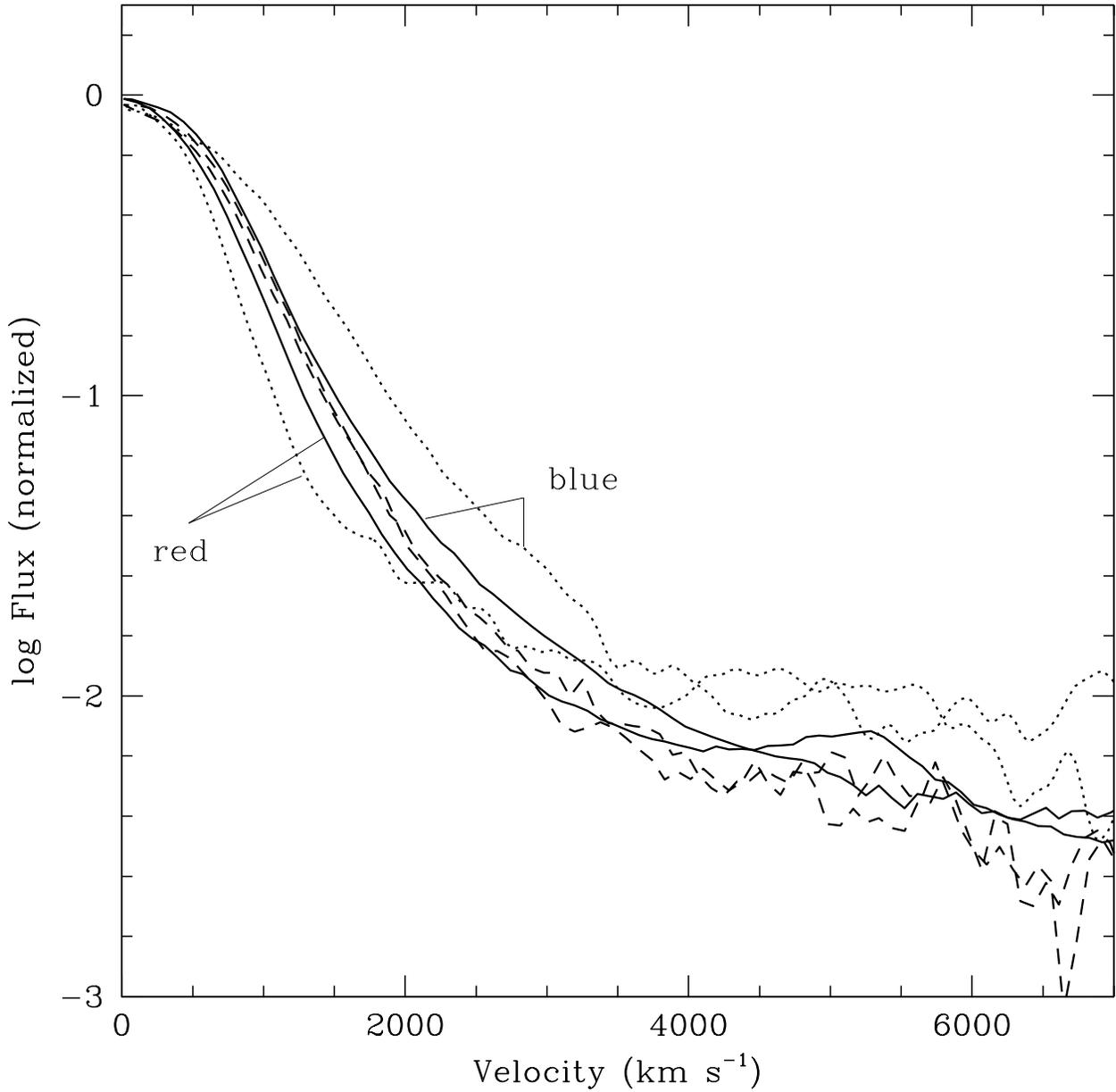}
\caption{Evolution of the H$\alpha$ line profile from 1996 (Lick; 
dashed line), to 1997 (Keck; solid line), to 1999 (VLT; dotted
line). Note the strong
decrease in the red wing of the line relative to the blue wing in the 1997
and 1999 spectra.}
\label{fighaevol}
\end{figure}

\clearpage

\begin{deluxetable}{lrlcc}
\tabletypesize{\scriptsize}
\tablecaption{Log of observations \label{tab1}}
\tablewidth{0pt}
\tablehead{
\colhead{UT Date}  & 
\colhead{Epoch\tablenotemark{a}}  & \colhead{Telescope}  &
\colhead{Wavelength Range} &\colhead{Resolution}\\
\colhead{}     &\colhead{days} &\colhead{} &
\colhead{(\AA)}&\colhead{(\AA)}}
\startdata 
24 May 1995 & 321 & Lick 3 m & 3120\phn -- \phn10490 & 6--11 \\
6--7 July 1995 & 364--365 & Lick 3 m & 3120\phn -- \phn10490 & 6--11 \\
1 August 1995 & 390 & Lick 3 m & 4260\phn -- \phn7040 & 6 \\ 
22 June 1996 &  716 & Lick 3 m & 3120\phn -- \phn9880 & 6 \\
6  February 1997 &  943 & HST & 1140\phn -- \phn4780 & 2--7 \\
11 April 1997 &  1007 & Keck I & 3910\phn -- \phn8880 & 10 \\
16 April 1997 & 1012 & Lick 3 m & 3160\phn -- \phn9870 & 6 \\
11 May  1997 &  1037 & Lick 3 m & 3120\phn -- \phn9880 & 6--11 \\
13 May  1999 & 1769 & VLT & 3635\phn -- 11680 & 13 \\
12 June  1999 & 1799 & VLT & 3908\phn -- \phn9187 & 5 \\
\enddata
\tablenotetext{a}{Based on an assumed explosion date of 4 July 1994.}
\end{deluxetable}

\begin{deluxetable}{lccrrrrl}
\tabletypesize{\scriptsize}
\tablecaption{Fluxes of narrow lines \label{tab2}}
\tablewidth{0pt}
\tablehead{
\colhead{Line}  & 
\colhead{Rest }  & \colhead{Observed}  &
\colhead{} &\colhead{} &\colhead{Flux\tablenotemark{a}} &\colhead{} \\
\colhead{}     &\colhead{wavelength} &\colhead{wavelength} &
\colhead{22 June}&\colhead{6 Feb}& \colhead{11 April}& \colhead{12 June}& \colhead{Notes} \\
\colhead{}     &\colhead{(\AA)} &\colhead{(\AA)} &
\colhead{1996} & \colhead{1997}& \colhead{1997}& \colhead{1999}& \colhead{}  }
\startdata 
 ~[Fe XI]          &   7891.8          &  7891.9       &         & & 1.2\phn &          0.12    &      \\
 ~     ?           &                   &  7610.3       &         & & 0.81    &          0.17    &      \\
 ~[Ar III]         &   7135.8          &  7135.7       &         & & 0.26    &          0.12    &  \\
 ~He I  (n)        &    7065.3         &    7065.5     &         & &         &          0.11    &  \\
 ~[Ar V]           &   7005.7          &  7005.0       &         & & 0.21    &                  &broad?  \\
 ~H$\alpha$        &   6562.8          &     6563.8    &         & &$\lsim$ 2.1\phn&          0.78&  \\
 ~[Fe X]           &   6374.5          &  6375.5       & 1.8\phn & & 1.3\phn  &         &  \\
 ~[O I]            &   6363.8          &    6363.0     &         & & 0.47    &                  & \\
 ~[O I]            &   6300.3          &    6301.0     & 3.0\phn & & 3.0\phn 
 & 0.22      & \\
 ~[Fe VII]         &   6087.0          &    6086.8     & 3.7\phn & & 2.8\phn & 0.60    &  \\
 ~[N II]           &   5754.6          &     5755.5    &         & & 2.2\phn &         &broad? \\
 ~[Fe VII]         &   5720.7          &     5721.9    & 3.0\phn & & 1.6\phn & 0.7\phn &  \\
 ~[Fe VI]          &   5677.0          &     5677.5    &         & & 0.50    &         &    \\
 ~[Fe VII]+[Fe VI] &   5276.4+5277.8   &      5278.1   & 6.5\phn & & 1.9\phn &         &bl. \\
 ~[Fe VI]          &   5176.0          &      5176.2   &         & & 3.3\phn & 0.11    &bl. w. Fe II, Fe VII  \\
 ~[Fe VII]         &   5158.9          &      5158.7   &         & &         & 0.20    &bl. w. Fe II, Fe VI  \\
 ~[Fe VI]          &   5145.8          &      5148.2   &         & &         & 0.19    &bl. w. Fe II, Fe VII  \\
  ~[O III]         &          5006.8   &      5006.7   & 44.\phn\phn&& 26.\phn\phn & 3.5\phn     &  \\
  ~[O III]         &          4959.0   &      4958.6   &          & & 8.5\phn     &          0.62    &  \\
 ~H$\beta$         &   4861.3          &     4861.6    &          & &$\lsim$3.6\phn&        0.51    &  \\
 ~[Ne IV]          &  4725.7           &     4725.4    &          &1.8\phn & 0.88            &         &  \\
 ~[Ne IV]          &  4714.2 - 4715.7    &     4714.8    &          & 0.43? & 0.41             &         &  \\
 ~He II            &    4686           &      4686.0   &          &1.5\phn & 0.97            & 0.11    &  \\
 ~[O III]          &  4363.2           &     4364.0    & 6.5\phn  & 8.3\phn & 6.4\phn          & 0.53    &  \\
 ~H$\gamma$        &   4340.5          &     4341.0    &          &  &         &          0.27    &  \\
 ~[Fe IV]         &  4198.2           &     4198.1    &          & & 0.07    &                 &  \\
 ~ ?               &                   &     4187.7    &          & & 0.05    &                  &  \\
 ~[Fe V]           &  4180.6           &     4179.8    &          & &2.1\phn     &                  &bl. w. He I 4169  \\
 ~[Fe V]           &  4071.2           &     4070.9    & 3.0\phn  & &1.3\phn     &          0.11     &  \\
 ~[Fe III]?        &  4046.2           &     4045.0    &          & &0.04   &                  &ident.?  \\
 ~He I            &  4009.3           &     4009.0    &          & &0.10   &                 &  \\
 ~[Ne III]         &  3967.5           &     3964.8   &          & 3.4\phn& &                  &bl. w. He I 3964.1\\
 ~[Ne III]         &  3868.8           &     3868.1    & 20.\phn\phn& 9.5\phn&         6.8\tablenotemark{b}\phn &         &  \\
 ~[Fe V]+[Fe VII]  &  3757.6+3758.9    &     3757.7   & 6.3\phn  & 3.4\phn &  9.0\tablenotemark{b}\phn &         &bl.\\
 ~[Fe VII]         &  3586.3           &     3585.3   &          & 2.1\phn & &                  &  \\
 ~[Ne V]           &  3425.9           &     3419.0   & 15.\phm{00}&&         &                  &  \\
 ~[Ne III]         &  3342.4           &     3343.6   &          & 3.8\phn &&                 &  \\
\enddata
\tablenotetext{a}{Observed fluxes in units of $10^{-16} \ergs {\rm cm}^{-2}$.}
\tablenotetext{b}{Actually measured in 11 May 1997 spectrum.}
\end{deluxetable}

\begin{deluxetable}{lccrrrrl}
\tabletypesize{\scriptsize}
\tablecaption{Fluxes of intermediate lines \label{tab3}}
\tablewidth{0pt}
\tablehead{
\colhead{Line}  & 
\colhead{Rest }  & \colhead{Observed}  &
\colhead{} &\colhead{}&\colhead{Flux\tablenotemark{a}} &\colhead{} \\
\colhead{}     &\colhead{wavelength} &\colhead{wavelength} &
\colhead{22 June}&\colhead{6 Feb}& \colhead{11 April}& \colhead{12 June}& \colhead{Notes} \\
\colhead{}     &\colhead{(\AA)} &\colhead{(\AA)} &
\colhead{1996} & \colhead{1997}& \colhead{1997}& \colhead{1999}& \colhead{}  }
\startdata
 ~O I              &  7771.9-7775.3    &  7773.5       & 1.4     & & 2.0     &          1.1     &  \\
 ~[O II]           &  7319.9, 7330.2    &  7326.5       & 5.3      & & 4.6       & 5.4     & bl. w. He I, Ca II\\
 ~[O I]            &   6300.3, 6363.8   &  6335.2       & 4.5      & & 3.2         & 4.4     &  \\
  ~[O III]         &   4959.0, 5006.8   &      5006.7   & 54.\phn  & & 39.\phn  & 9.5     &  \\
 ~Mg I]            &    4571.1         &   4574.1      & 1.8      &1.3 & 1.8         & 1.6     & (+4583) \\
 ~[O III]          &  4363.2           &     4380.7    &          &5.3 & 4.7             & 1.2     &bl. w. H$\beta$  \\
 ~Mg I             &   2852.1     &    2862.3     &          & 2.5 & &               &bl. w. Mg II \& Fe II\\

 ~Si III], C III]   &1882.7, 1892.0, 1906.7, 1909.6&    1899.3     &         &  21.\phn     &        &         &bl.    \\
 ~N III]           &1746.8 - 1753.4    &    1752.9     &          &
 2.1     & &                &    \\
 ~O III]           &    1660.8, 1666.1        &    1658.8     &          & 9.9     & &                &   \\
 ~C IV             &  1548.9, 1550.8   &    1548.0     &          & 7.6     & &                 &   \\
 ~N IV]           &1483.3-1487.9 &    1492.5     &          & 2.2     & &              &flux uncertain\\
 ~Si IV, O IV]     &1393.8, 1402.8, 1397.2 - 1407.4&    1406.3     & & 6.7     &         &         &bl.     \\
 ~Si II, O I       &1304.4, 1309.3, 1302.2 - 1304.9&    1310.0     &          & 2.1     &        &         &ident. uncertain\\
  &\\

\enddata
\tablenotetext{a}{Observed fluxes are in units of $10^{-15} \ergs {\rm cm}^{-2}$.}
\end{deluxetable}

\clearpage

\begin{deluxetable}{lccrrrrl}
\tabletypesize{\scriptsize}
\tablecaption{Fluxes of broad lines \label{tab4}}
\tablewidth{0pt}
\tablehead{
\colhead{Line}  & 
\colhead{Rest }  & \colhead{Observed}  &
\colhead{} &\colhead{}&\colhead{Flux\tablenotemark{a}} &\colhead{} \\
\colhead{}     &\colhead{wavelength} &\colhead{wavelength} &
\colhead{22 June}&\colhead{6 Feb}& \colhead{11 April}& \colhead{12 June}& \colhead{Notes} \\
\colhead{}     &\colhead{(\AA)} &\colhead{(\AA)} &
\colhead{1996} & \colhead{1997}& \colhead{1997}& \colhead{1999}\tablenotemark{b}& \colhead{}  }
\startdata
 ~He I             &   10830           &  10827.6      &          &&         &          6.0\phn     &  \\
 ~P$\epsilon$, He I,[S III]?&   9545.7, 9529.3, 9532.0&   9544.0    &  3.0\phn &&                 &  0.70  &  \\  
 ~P$\zeta$         &   9229            &               &          &&         &               &  \\
 ~Fe II            &  9196.9-9218.2&           &          &&         &          0.18    &bl.  \\
 ~Fe II+He I       &  9175.9, 9078.1, 9174.5& 9177.5      & 7.0\phn       &         &         &  1.0\phn&bl.  \\
 ~Fe II            &  9122.9, 9132.4      &   9133.8      & 2.0\phn       &         &         &  0.39   &bl.  \\
 ~Fe II            &  9070.5, 9077.4 &   9071        & 0.54     &&                  &   0.07  &bl.  \\
 ~P$\eta$          &  9014             &               &          &         &         &  0.07   &bl.  \\
 ~Ca II            &  8662.1           &   8668.2      &         && 0.27    &          &bl. H I, Ca II  \\
 ~H I              &  8598.3           &  8601.7       &          && 0.07   &                  &bl.  \\     
 ~H I, Ca II       &  8545.4+8542.1    &  8545.9       & 0.20     &&0.22     &                  &bl.  \\     
 ~Ca II            &  8498.0           &  8496.9       &          &&0.80\phn     &                  &bl.   \\
 ~O I + Fe II      &  8446.5 + 8451.0  &  8448.1       & 8.4\phn  && 3.5\phn     & 0.88     &bl.  \\
 ~He I  	   &  8285.4           &  8287.3       & 0.16      && 0.12        & 0.05   &  \\
 ~Fe II 	   &  8228.9           &  8227.7       & 0.72      && 0.53             & 0.11     &  \\
 ~He I  	   &  8155.8           &  8156.3       &          && 0.02             & 0.02     &  \\ 
 ~He I  	   &  7971.6           &  7970.7       & 0.08      && 0.22          & 0.05     &  \\
  ?               &                   &  7892.0       &          && 0.14     &         &narrow?    \\
  ?               &                   &  7610.4       &          && 0.07     &         &narrow?    \\
 ~He I             &    7281.0         &   7277.7      &          && 0.46             &         &bl. w. [O II]  \\ 
 ~Fe II            & 7155.1, 7172.0     &  7156.7       &          && 0.21              &         &mult. 14  \\
 ~He I             &    7065.3         &    7063.5     & 2.0\phn       && 1.2\phn        & 0.40     &  \\
 ~He I             &   6678.2          &  6675.6       & 0.60      && 0.33          &         &  \\
 ~H$\alpha$        &   6562.8          &     6562.4    & 330.\phn\phn && 160.\phn\phn       & 30.\phn\phn     &  \\
 ~[O I]            &   6300.3, 6363.8   &  6335.2       & 4.5\phn      && 3.2\phn         & 4.4\phn     &  \\
 ~Fe II            &   6148, 6149       &  6146.3       &              && 0.19                     &&mult. 74   \\
 ~He I             &   5875.7          &     5876.4    & 3.8\phn      && 2.5\phn         & 0.88     &  \\\
 ~[N II]           &   5754.6          &     5753.5    & 0.39      && 0.27     &          0.11     &  \\
  ~Fe II           &   5534.9          &     5533.2    &          && 0.13     &                   &mult. 55, bl. w. [O I] 5577 \\
 ~Fe II            &   5414.1, 5425.3   &     5421.1    &          && 0.18     &                  &mult. 48+49\\
 ~Fe II            &   5362.9          &     5367.1    &          && 0.22     &                  &mult. 48  \\
 ~Fe II            &   5316.2          &     5317.9    &          && 0.54     &                  &mult. 49  \\
 ~Fe II            &   5265, 5276.0     &     5273.9    &         && 0.10     &                  &mult. 48+49  \\
  ~Fe II           &   5169.0          &     5169.5    & 2.6\phn  && 1.0\phn         &         &mult. 42, bl. w. [Fe VI], [Fe VII] \\
  ~Fe II           &   4629.3          &     4627.6    &          && 0.25                     &         &mult. 37 \\
 ~H$\beta$         &   4861.3          &     4861.6    & 8.8\phn      && 4.3\phn         & 1.2\phn     &  \\
 ~He II            &    4685.7         &      4686.0   & 0.77      &0.24& 0.26         &         &  \\
 ~He I             &   4471.5          &    4470.7     & 1.1\phn      &0.63& 0.68             & 0.26     &  \\
 ~[O III]          &  4363.2           &     4380.7    &          &5.4\phn& 4.7\phn              & 1.2\phn     &bl. w. H$\beta$  \\
 ~H$\gamma$        &   4340.5          &     4341.0    &          &4.6\phn& 3.4\phn              & 1.1\phn     &bl. w. [O III]  \\
  ~Fe II           &   4233.2            &     4238.9    &          && 0.48     &                  &mult. 27\\
  ~Fe II           &   4178.9            &     4180.3    &          && 0.22     &                  &mult. 28, bl. w. [Fe V] 4180\\
 ~He I             &   4120.8          &    4122.4     &          &&    ?    &             &  \\
 ~H$\delta$        &  4101.7             &     4102.3    & 1.7\phn      && 1.6\phn              & 0.78     &  \\
 ~[Ne III], H$\epsilon$         &  3967.5, 3970.0             &     3967.8    & 2.2\phn      && 0.93        & 0.30        &blend w. Fe V, H$\epsilon$  \\
 ~H$\zeta$         &  3889.0             &               &          &&         &                 &  \\
 ~He I             &  3888.6           &               & 1.5\phn      &&         &                 &  \\ 
 ~H$\eta$          &  3835.4             &               & 1.2\phn      &1.1\phn&                  &         &  \\
 ~[Ne V]           &    3426           &       3447    & 1.8\phn      &&         &                  &  \\  
   ~?               &   ? ,         &     3075.4   &          &0.58&      &                 &uncert. id.\\
   ~?               &   ? ,         &     3001.1   &          &0.35&      &                  &uncert. id.\\
   ~O III           &   ? ,         &     2983.5   &          &0.36&      &                  &uncert. id.\\ 
  ~Fe II, [Mg V]    &   2926.6 , 2928.3  &     2925.8   & &0.23&      &                &mult. 60, ident. uncert.\\
 ~Mg II            &   2795.5, 2802.7  &    2793.3     &          & 130.\phn\phn      &         &         &   \\   
  ~He I + Al II]   &   2663.4+2669.2       &     2666.3    &          &0.71&             &         &\\
  ~Fe II           &   2614.9-2631.0&     2617.7   &          &3.9\phn&      &                  &mult. 1+171\\
  ~Fe II           &   2585.9 -2593.7  &     2592.0    &          &0.85&      &                  &mult. 1+64\\
  ~Fe II           &   2562.5, 2563.5  &     2561.3    &          &0.35&      &                  &mult. 64\\
  ~Fe II           &   2506.4-2508.3   &     2505.6    &          &2.0\phn&      &                  &\\
 ~Ly$\alpha$       &    1215           &    1210.9     &          & 300.\phn\phn &                 &     \\

  
\enddata
\tablenotetext{a}{Observed fluxes are in units of $10^{-15} \ergs {\rm
cm}^{-2}$.}
\tablenotetext{b}{Includes some measurements from the 13 May 1999 VLT spectrum,
especially in the near-IR.}.
\end{deluxetable}

\clearpage

\end{document}